\documentclass[12pt,aaas,science]{revtex4-1}

\usepackage{multirow}
\usepackage[utf8]{inputenc}
\usepackage{graphicx}
\usepackage{subfigure}
\usepackage[usenames,dvipsnames]{xcolor}
\usepackage{epstopdf}
\usepackage{amsmath,amsthm,amssymb}
\usepackage{latexsym}
\usepackage{bm}
\usepackage{dcolumn}
\usepackage{braket}
\usepackage[normalem]{ulem}
\usepackage{times}
\usepackage{hyperref}

\newcommand{\ignore}[1]{}
\newcommand{\boxedpp}{\boxed{{}_+^+}}
\newcommand{\boxedpm}{\boxed{{}_+^-}}
\newcommand{\boxedmp}{\boxed{{}_-^+}}
\newcommand{\boxedmm}{\boxed{{}_-^-}}
\newcommand{\boxedpppp}{\boxed{{}_+^+{}_+^+}}
\newcommand{\boxedpppm}{\boxed{{}_+^+{}_+^-}}
\newcommand{\boxedppmp}{\boxed{{}_+^+{}_-^+}}
\newcommand{\boxedppmm}{\boxed{{}_+^+{}_-^-}}
\newcommand{\boxedpmpm}{\boxed{{}_+^-{}_+^-}}
\newcommand{\boxedpmmp}{\boxed{{}_+^-{}_-^+}}
\newcommand{\boxedpmpp}{\boxed{{}_+^-{}_+^+}}
\newcommand{\boxedmppp}{\boxed{{}_-^+{}_+^+}}

\begin{document}
\bibliographystyle{naturemag}

\title{
Finding and classifying an infinite number of cases of the marginal phase transition in one-dimensional Ising models}
\author{Weiguo Yin}
\email{To whom correspondence should be addressed; E-mail: wyin@bnl.gov.}
\affiliation{Condensed Matter Physics and Materials Science Division,
Brookhaven National Laboratory, Upton, New York 11973, USA}

\begin{abstract}
\textbf{
One-dimensional systems---ranging from travelling light to circuit cables and from DNA to superstrings---are ubiquitous and critically important to the human knowledge of the universe. However, our engagement with one-dimensional systems in the research and education of spontaneous phase transitions, the phenomena wherein materials can change rapidly between different phases (e.g., gas, liquid, solid, etc.) on their own, has not been largely exercised, since it was proven that one-dimensional systems do not contain phase transitions in the textbook Ising model almost 100 years ago~\cite{Ising1925} and its quantum counterpart, the Heisenberg model, over 50 years ago~\cite{Mermin_PRL_theorem}. Recently, a spontaneous marginal phase transition (MPT) was discovered in a one-dimensional Ising model containing strong geometrical frustration~\cite{Yin_MPT}. Here, by exploring the symmetry of the new mathematical structure underlying the MPT, I report the finding and classification of an infinite number of MPT cases---with highly tunable intriguing behaviors like phase reentrance, the dome shape of transition temperature, pairing, and gauge freedom. These discoveries reveal the possibility of building the MPT-based one-dimensional Ising Machine that can be used to simulate the complex phenomena of phase competition in strongly correlated systems and provide insights with its unambiguous exact solutions. They also form a rich playground for exploring unconventional phase transitions in  one-dimensional Heisenberg models.}

\ignore{open the door to new interdisciplinary researches in
studying, engineering, and utilizing the rich phenomenology of MPT, ranging from building the first-generation phase-transition-ready one-dimensional circuits to developing advanced materials theory for frustration-driven phase transitions in strongly correlated electronic materials.}

\ignore{The new idea was inspired by mapping the original spin system to a social-science issue with increasing family sizes, in which the children's actions create the ``frustration function,'' the math core of the diversity in MPT.}

\end{abstract}

{
}
\date{\today}

\maketitle


\section{Introduction}

During the past 30+ years, monumental experimental and theoretical efforts have been made to understand the complexity in strongly correlated electronic systems \cite{Dagotto_Science_review} such as copper-oxide high-temperature superconductors, manganese oxides with colossal magnetoresistance, iron-based high-temperature superconductors, iridium-based relativistic Mott insulators, etc. A hallmark of this complexity is the existence of many competing phases that have distinct symmetries, yielding spontaneous nanoscale electronic inhomogeneity~\cite{Moreo_Science_1999,Pan_Nature_Davis,Tranquada2004}. The phenomenology of phase competition has promising consequences for energy and information technology applications, because it implies giant responses to tiny external stimuli. The underlying driving force for this complexity is the dynamic interplay of the charge, spin, orbital, and lattice degrees of freedom~\cite{Moreo_Science_1999,Yin_PRL_LaMnO3,Yin_PRL_FeTe}. Simply for geometrical reasons, it is usually already difficult to satisfy all of those degrees of freedom and their interactions,
a phenomenon called \emph{geometrical frustration}~\cite{Edwards_1975_Anderson,2011_book_frustration,Balents_nature_frustration}. When no single order can dominate the competition in a frustrated magnet, a quantum spin liquid state emerges---with strong entanglement between long distant spins, a feature useful for quantum information~\cite{Balents_nature_frustration,Yin_PRL_pyroxene}. While geometrical frustration is commonly known to suppress ordering, it was once envisioned to drive phase transition to superconductivity at low temperature due to the necessity to relieve the entropy accumulated by frustration~\cite{Si_PRL_FeSC}; however, precisely how this could happen remains elusive. Moreover, even in the absence of geometrical frustration, there exists exotic \emph{quantum frustration}, namely the fierce battle among the different phases (or the different degrees of freedom) for the interactions on each single atom-atom bond. For example, in doped copper oxides, the spins try to utilize the superexchange interactions to realize the antiferromagnetic order, but only find that exactly the same interactions are being used by charge carriers to party in the superconductivity~\cite{Lee_PRB_Ubbens}. The electrons, the composite entities of the charge and spin degrees of freedom, are strongly frustrated in this real-space battle, and as a way out, they transform into distinguished nodal and antinodal states in the momentum space~\cite{Valla_Science_99,Hashimoto_NP_ZXShen} or  resort to a world of higher dimension for  solutions~\cite{Zhang_SO5}. Alternatively, a compromise may be reached to allow real-space domains, stripes, or clusters of different orders, yielding the spontaneous nanoscale electronic inhomogeneity~\cite{Moreo_Science_1999,Pan_Nature_Davis,Tranquada2004}. Precisely how the electrons deal with strong frustration, either quantum or geometrical, and establish the complex phase diagram is one of the challenging questions in strongly correlated electron systems.

\ignore{Note that quantum frustration describing an individual electron or atom's feeling about the macroscopic phase competition is coined here as a complement to the common terminology of quantum fluctuation, which implies a correlated behavior around a macroscopic ordered phase. Anyhow,}

The challenge stems from the fact that frustration is in our  knowledge domain far away from the around-mean-field treatment, such as Ginzburg-Landau plus mean-field theory~\cite{Chakravarty_Nature_04,Yin_PRB_phaseCompetition,Hirschfeld_phaseCompetition}. Developing accurate theoretical or computational methods for thermodynamic simulation of strongly correlated and frustrated electronic systems, especially for the phase transition problems that often require the reach of the thermodynamic limit (i.e., infinite size), is one of the most challenging questions~\cite{2011_book_frustration,2004_book_quantum_magnetism,tsvelik_book_2003}. For example, quantum Monte Carlo would likely suffer from the sign problem for strongly frustrated systems~\cite{White_J1_J2,Maisinger_PRL_J1-J2_TDMRG,Feiguin_PRB_TDMRG}. As such, despite the significant achievements in understanding the physical effects of frustration~\cite{2011_book_frustration,2004_book_quantum_magnetism,tsvelik_book_2003,Balents_nature_frustration,Balents_NP_07_frustration,Chandra_frustrated_Heisenberg_PRL_90_order-by-disorder,Henley_PRL_89_order-by-disorder,Shender_1982_order-by-disorder,White_J1_J2,Edwards_1975_Anderson,Anderson_CuMn,Fazekas1974_Anderson,Kitaev_2006}, relatively little is known about the behavior of frustrated quantum spin systems, in comparison with their unfrustrated counterparts. Whereas, on the experimental side, with recent remarkable advances in photon sciences, especially in ultrafast photoexcited phase transitions~\cite{Yin_npjQM_2016} and in x-ray total scattering analysis that can provide structural information from high-temperature ``disordered'' materials~\cite{Bozin_NC_CuIr2S4,Yin_PRL_nicklate}, more and more pieces of evidence about the peculiarities of the intermediate-temperature regime have been demonstrated. For example, the Bozin-Billinge orbital degenerate lifting (ODL) states, which are not the around-mean-field fluctuations of the low-temperature orders, were discovered in a number of frustrated materials~\cite{Bozin_NC_CuIr2S4}, while Cao \textsl{et al.} found a strange frustration and decoupling of spin chains in the trimer magnet Ba$_4$Ir$_3$O$_{10}$~\cite{Cao_npjQM_Ba4Ir3O10}, reminiscent of the frustration-driven decoupling of spin chains in the Heisenberg model in a zigzag ladder~\cite{White_J1_J2} and in a two-dimension lattice~\cite{Balents_NP_07_frustration}.
These new data motivated a theoretical study of the one-dimensional Ising model on a 2-leg ladder~\cite{Dagotto_science_ladder}---but with novel trimer rungs~\cite{Yin_MPT}. This model was exactly solved to show precisely how strong geometrical frustration effects decouple the chains in a non-mean-field way, resulting in an exotic ``normal state''~\cite{Yin_MPT}.

Unexpectedly, a super sharp transition at a lower temperature was recorded in this study of the Ising model in one dimension~\cite{Yin_MPT}. While it was exciting to see the transition reminiscent of the speculation that the relief of entropy accumulated by frustration can take the form of creating another order~\cite{Si_PRL_FeSC}, it was even more astonishing considering that spontaneous phase transitions at any nonzero are forbidden in the one-dimensional Ising models with short-range interactions~\cite{Ising1925,Mattis_book_1985,Baxter_book_Ising,Cuesta_1D_PT}. This transition has a large latent heat of $T_{m}\ln2$ (where $T_{m}$ is the transition temperature) and can have a transition width of less than $10^{-30}$ K; so, it would be characterized as a phase transition in lab heat capacity measurements. This transition features an unconventional order parameter (instead of magnetization, the conventional type) that describes the long-range orders of a intra-rung two-spin correlation function and are classified by whether the said correlation function is ferromagnetic or antiferromagnetic. The transition at $T_{m}$ is the switching between these two ``ferromagnetic'' or ``antiferromagnetic'' orders (they have zero magnetization). When the transition width becomes absolute zero, $T_{m}=0$ as well, so it does not violate the theorems~\cite{Cuesta_1D_PT}. The transition is thus referred to as a marginal phase transition (MPT)~\cite{Yin_MPT}. The spontaneous MPT exhibits exceptional beauty: it contains a new mathematical structure that resembles the original, simplest one-dimensional Ising model~\cite{Ising1925} in the presence of an external field---and with the external field replaced by a purely on-rung contribution of the trimers. It is this internal-field-like frustration effect that drives the MPT.

These observations point to the possibility of using the MPT-based one-dimensional Ising model to simulate the complex phase competition in strongly correlated systems, i.e., building an \emph{Ising Machine}. To this end, highly tunable parameters for various design and control are needed. Therefore, the fate of MPT in this research direction is determined by whether this is a very special case or not. In retrospect, one instance of MPT in an Ising tetrahedral chain composed of edge-sharing tetrahedra was reported in the literature~\cite{Strecka_Ising_pseudo}; however, without the knowledge of its unconventional order parameter, this instance was categorized as a pseudo phase transition alongside many decorated single-chain Ising models in the presence of an external magnetic field~\cite{Souza_SSC_18_pPT,Rojas_PRE_99_pPT,Derzhko_pPT} and their qualitative difference has been completely overlooked [see text around Eq.~(\ref{pPT}) for more details]. In terms of the degree of locality in the order parameter, MPT is only one step away from conventional phase transitions; thus, it is likely to be generalized. Here, inspired to explore the supersymmetry~\cite{Zhang_SO5} of the mathematical structure of the MPT, I report the finding and classification of an infinite number of cases of MPT---with highly tunable intriguing behaviors like phase reentrance, $T_{m}$ domes, pairing, gauge freedom, etc. The existences of such countless MPT cases and exotic phases in the exactly solved Ising model are stimulating. They are anticipated to stimulate the simultaneous development of phase-transition-ready one-dimensional devices and advanced materials theory for frustration-driven phase transitions. The present results mark a start toward realizing a MPT-based one-dimensional Ising Machine, which will be used to simulate the complexity in arbitrary strongly correlated systems and provide insights with its unambiguous exact solutions. In addition, they form a rich playground for exploring phase transitions in one-dimensional quantum Heisenberg models, in which long-range orders of the conventional type are forbidden as well~\cite{Mermin_PRL_theorem,Hohenberg_PR_67_noPT,Coleman_1973_noPT}. Since the additional strong quantum fluctuations in one dimension tend to prevent phase transitions at finite temperature, a lack of the one-to-one classical-quantum correspondence of the phase transitions is expected. Therefore, it is particularly constructive to have a pool of an infinite number of MPT cases to start with in the search for unconventional phase transitions in one-dimensional Heisenberg models,

This article is organized as follows: Section II introduces a new parent-children version of the Ising model to facilitate the presentation. Section III describes how to exactly solve the model and find MPT on the basis of symmetry analysis with the math presented at the high-school pre-calculus level. Section IV classifies the resulting infinite number of cases of MPT. Section V shows and discusses examples in each class with exotic properties. Section VI presents further symmetry-based discussions on gauge freedom and extensions of the model to include decorations in between the households and to develop the supercell method. Section VII summarizes the results, focusing on their implications on achieving the MPT-based one-dimensional Ising Machine.

\section{The Ising model as a social system}

\begin{figure}[t]
    \begin{center}
        \subfigure[]{
\includegraphics[width=\columnwidth,clip=true,angle=0]{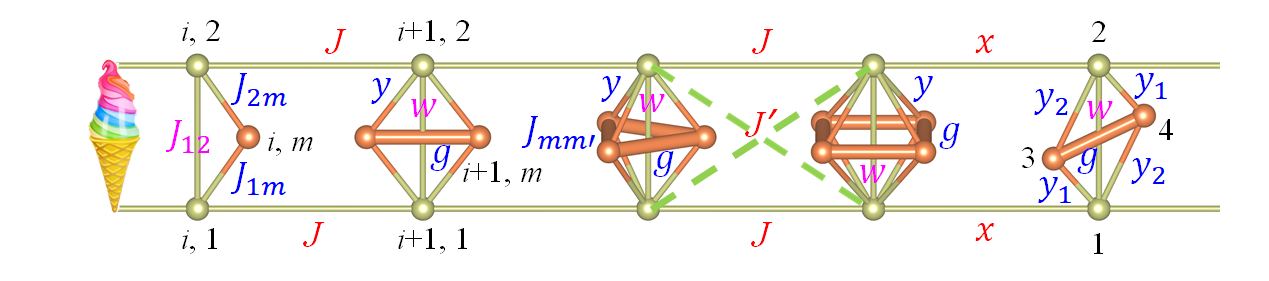}
        }
\vspace*{0.2in}
\newline\newline
        \subfigure[][]{
\includegraphics[width=0.48\columnwidth,clip=true,angle=0]{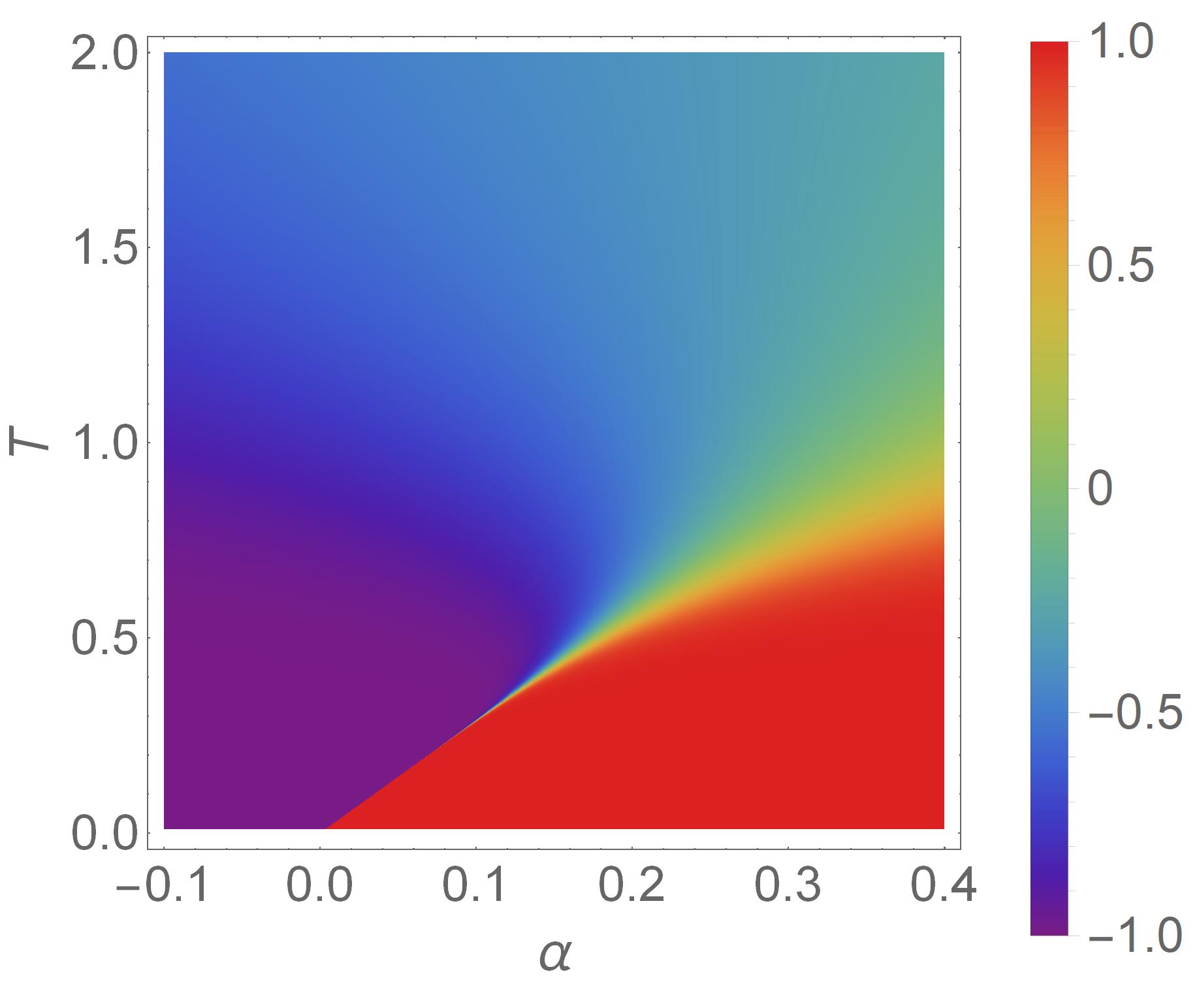}
        }
        \subfigure[][]{
\includegraphics[width=0.48\columnwidth,clip=true,angle=0]{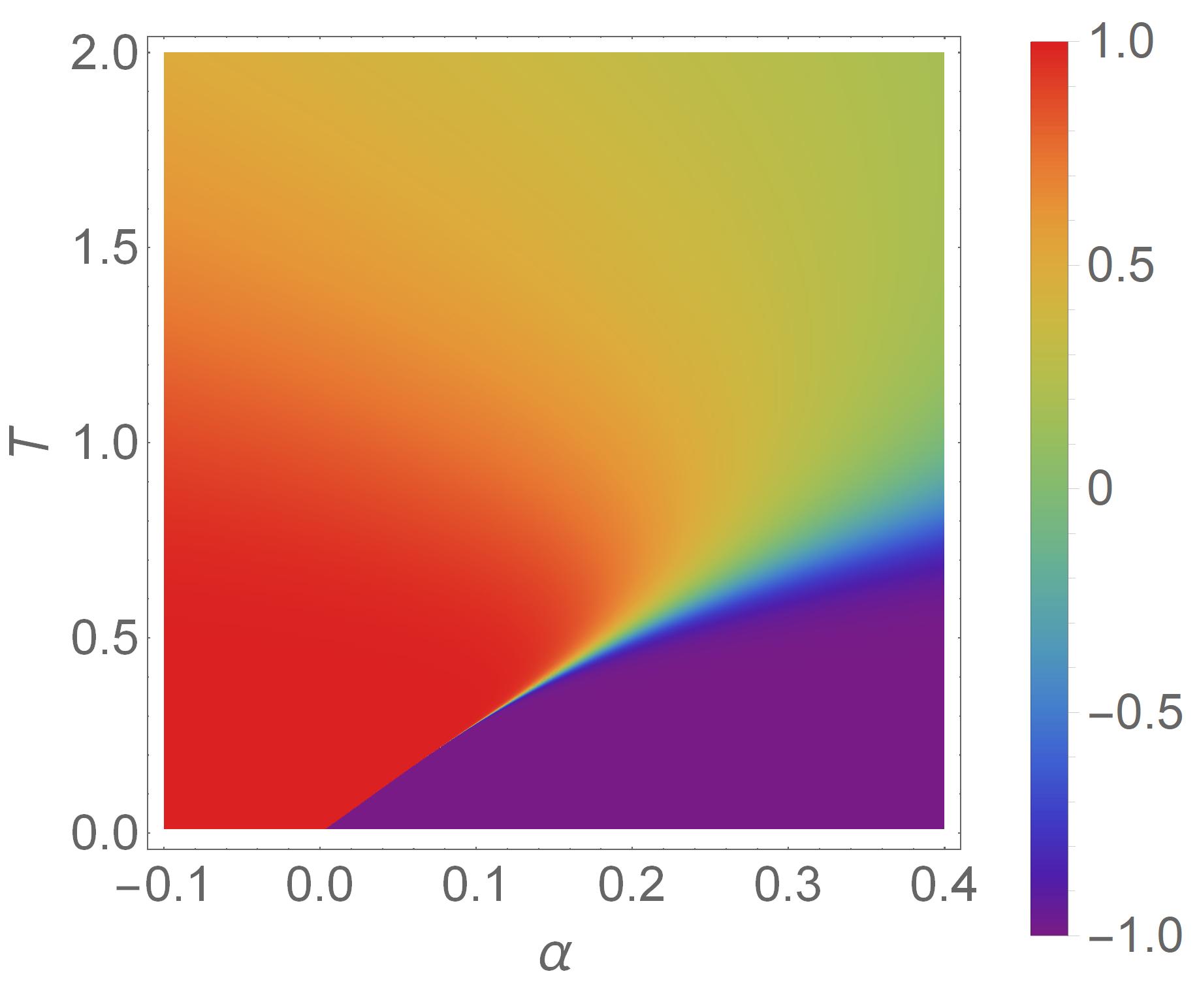}
   }
   \end{center}
\caption{\textbf{The model.} (a) The 2-leg ladder with (from left to right) ice-cream-cone, triangle, diamond, trigonal-bipyramid, octahedron, and exotic diamond rungs. The outer (yellow) and inner (orange) balls stand for the parent and children spins, respectively. The bonds represent the interactions $J (x)$, $J'$ (dashed lines), $J_{12} (w)$, $J_{1m}/J_{2m}$ ($y$ or $y_1, y_2$), and $J_{mm'} (g)$: the letters in the parentheses are shorthand notations for their values multiplied by the inverse temperature $\beta=1/T$. Unlike the other regular rungs, which have the mirror symmetry to link the two parents, the deformed diamond rung (rightmost) has the inversion symmetry but no mirror symmetry. \textbf{The phase diagrams of the model with the deformed diamond rungs} in terms of the order parameter $C_{12}(0)$ as a function of the temperature $T$ and the frustration parameter (b) $\alpha=(|y_1+y_2|+w)/|x|$ for the regular case $g=3$ and  (c) $\alpha=(|y_1-y_2|-w)/|x|$ for the exotic case $g=-3$. Red stands for the $+1$ region (where the parents have like values), purple stands for the $-1$ region (where the parents have unlike values). The sharp transitions between them with $T_{m}=2\alpha|J|/k_\mathrm{B}\ln2$ take place for strong frustration $0<\alpha < 0.1$. $J/k_\mathrm{B}$ is the temperature unit for $T$.}
\label{Fig:model}
\end{figure}

The generalized one-dimensional model is depicted in Fig.~\ref{Fig:model}a. It is a ladder with advanced rungs: the spins on each rung form a polyhedron (or an ice cream cone to symbolize the far richer details the rung can have). It looks unnatural and not easy to imagine. However, once the model is mapped to a social system, everyone would agree that it makes sense. Suppose this model stands for a neighborhood on a long street. Every rung corresponds to a household. The two spins located on the legs are the two parents.
The parents talk to the parents of the nearest neighbors. In the original model~\cite{Yin_MPT}, every household has only one child, who makes essentially a triangle with his/her argumentative parents. Now it is clear that introducing  polyhedral rungs means increasing the family size. \ignore{This model could be representative of United States, as the average American family has four members, for example.} We hold the following mapping convention: ladder=street, rung=household, outer spins=parents, and inner spins (or other subjects)=children. The language in terms of parents and children makes the presentation easier.

Expressed in math, one of the infinite possible forms of the generalized 2-leg ladder model with ice-cream-cone rungs is the following Ising model $H=\sum_{i=1}^{N}{\left[H^{(i,i+1)}_\mathrm{parents}+H^{(i)}_\mathrm{children}+H^{(i)}_\mathrm{bias}\right]}$ (where $N$ is the total number of the households and we are interested in the large $N$ limit) given by (c.f., Fig.~\ref{Fig:model}a)
\begin{eqnarray}
H^{(i,i+1)}_\mathrm{parents}&=&-J(\sigma_{i,1}\sigma_{i+1,1}+\sigma_{i,2}\sigma_{i+1,2})-J'(\sigma_{i,1}\sigma_{i+1,2}+\sigma_{i,2}\sigma_{i+1,1}) \nonumber\\
&& -\frac{1}{2}J_{12}(\sigma_{i,1}\sigma_{i,2}+\sigma_{i+1,1}\sigma_{i+1,2}),\nonumber \\
H^{(i)}_\mathrm{children}&=&-\sum_{m=3}^{M+2}\left(J_{1m}\sigma_{i,1}\sigma_{i,m}+J_{2m}\sigma_{i,2}\sigma_{i,m}\right) - \frac{1}{2}\sum_{m=3}^{M+2}\sum_{m'=3}^{M+2} J_{mm'}\sigma_{i,m}\sigma_{i,m'}, \label{model}
\\
H^{(i)}_\mathrm{bias} &=& -B {\left(g_1\sigma_{i,1}+g_2\sigma_{i,2}+\sum_{m=3}^{M+2}g_m\sigma_{i,m}\right)}, \nonumber
\end{eqnarray}
where $\sigma_{i,1}=\pm 1$ and $\sigma_{i,2}=\pm 1$ denote the two parents on the $i$th household (rung) of the street (ladder). Starting from the index number 3, $\sigma_{i,m}=\pm 1$ stands for the children on the $i$th household. $\sigma_{N+1,m}\equiv\sigma_{1,m}$ (i.e., the periodic boundary condition). $M$ is the number of children per household, which is an arbitrary natural number. There are infinite possibilities of the children distribution on the rungs.
$J$ and $J'$ are the interactions between parents of neighboring households. Inside one household, $J_{12}$ is the interaction between the two parents, $J_{1m}$ and $J_{2m}$ the interaction between the children and their parents, $J_{mm'}$ the interaction between child $m$ and child $m'$. To be complete, $B$ is the external bias field, which will influence the individuals' opinions or behaviors with a specific preference; it is irrelevant in this paper, as we are interested in spontaneous phase transitions at $B=0$.

As proved exactly in the next section, we stress here that the forms of interactions among the children can be not only two-body---as $J_{mm'}$ explicitly written in Eq.~(\ref{model})---but also three-body, four-body, $\ldots$, arbitrary-body interactions. Moreover, particularly for the field of quantum physics, quantum computing, quantum information, etc., the children spins can be quantum spins and the interactions among the children can be of quantum nature, too, with transverse components, or even not spins---because the commutator $[H^{(i)}_\mathrm{children},H]=0$---as long as the children-parents interactions are of the Ising type, i.e., the parents provide classical fields to their children. This abundance is exactly the reason that the fusiform rungs are called ice-cream-cone rungs, which include and are not limit to polyhedron rungs. The only two constraints on my proof are the following: (i) the parents are classical spins and they interact with their children using only classical interactions like the ones explicitly written in Eq.~(\ref{model}), no matter whether the children are classical or quantum. (ii) The system satisfies the \emph{weak condition} of the minimum unit cell size ($L=1$), meaning that the household dependences of $M$, member characters, and interactions are neglected; yet, the system has an infinite number of symmetry-related redundancy (or a gauge freedom), which can facilitate the optimization of the children's spatial distribution (see Section VI). It is noteworthy that this kind of lattice is generally called a decorated  lattice~\cite
{Fisher_decorated_59,Syozi_Ising_2D,Strecka_decorated_10,Rojas_decorated_11,Proshkin_decorated_19};
however, the studies of the present decorated 2-leg ladder model, Eq.~(\ref{model}), have not been found in the literature. The issue of further decorating the system in the inter-neighbor area is solved exactly in Section VI. This leads to an immediate application to addressing the important issues on the household dependences of the model parameters in terms of supercells with $L>1$, thus partially lifting constraint (ii).

\section{The exact solutions}
Our task is to compute the partition function $Z=\mathrm{Tr}\left( e^{-\beta H} \right)$, i.e.,
\begin{equation}
Z=\sum_{\{\sigma_{i,m}=\pm1\}}{e^{-\beta\sum_{i=1}^{N}{\left[H^{(i,i+1)}_\mathrm{parents}+H^{(i)}_\mathrm{children}+H^{(i)}_\mathrm{bias}\right]}}}
\label{partition}
\end{equation}
for all possible combinations of the values $\sigma_{i,m}$ ($i=1,2,3,\ldots,N$ and $m=1,2,3,\ldots,M+2$), i.e., $2^{N(M+2)}$ configurations, in the large $N$ limit. Here $\beta=1/(k_\mathrm{B}T)$ with $T$ being the temperature and $k_\mathrm{B}$ the Boltzmann constant. The thermodynamical properties are retrieved from the free energy per household $f(T)=\lim_{N\to\infty}-\frac{1}{N} k_\mathrm{B}T\ln Z$, the entropy $S=-\partial f/\partial T$, and the specific heat $C_v=T\partial S/\partial T$. The order parameter for MPT is the correlation function between the two parents: $C_{12}(0)=-\partial f/\partial J_{12}$~\cite{Yin_MPT}.

Eq.~(\ref{partition}), the summation over an infinite number of items, can be obtained exactly by using the transfer matrix method~\cite{Mattis_book_1985,Baxter_book_Ising,Yin_MPT}. (For one dimension, the method could be mastered by a precalculus-level high-school student.) In short, the partition function reads
\begin{equation}
Z=\mathrm{Tr}\left(\Lambda^N\right)=\sum_k{\lambda_k^N} \;\;\;\;\to\;\;\;\; \lambda^N \;\mathrm{for}\; N\to \infty,
\end{equation}
where $\Lambda$ is the transfer matrix, $\lambda_k$ the $k$th eigenvalue of $\Lambda$, and $\lambda$ the largest eigenvalue. So, the infinity of $N$ actually simplifies the answer. As a result, the free energy per household $f(T)=-k_\mathrm{B}T\ln\lambda$.

In constructing the transfer matrix $\Lambda$, we deal with the following object for the two neighboring households $i$ and $i+1$ in the absence of the external bias field:
\begin{equation}
 e_{}^{-\beta H^{(i,i+1)}_\mathrm{parents}-\frac{1}{2}\beta \left[H^{(i)}_\mathrm{children}+H^{(i+1)}_\mathrm{children}\right]}.
\label{transfer}
\end{equation}
The children's values can be exactly integrated out, as they interact only with the members of the same household and interact via the Ising-type interaction  with the parents of Ising type, which yields the following $4\times 4$ transfer matrix in the order of the two parents' values\\ $\left({}_{\sigma_1}^{\sigma_2}\right) = \left({}_+^+\right), \left({}_+^-\right), \left({}_-^+\right), \left({}_-^-\right)$:
\begin{eqnarray}
\Lambda=\left(
\begin{array}{cccc}
 e^{2x+2x'+w} {\boxedpp}_i{\boxedpp}_j & {\boxedpp}_i{\boxedpm}_j & {\boxedpp}_i{\boxedmp}_j & e^{-2x-2x'+w} {\boxedpp}_i{\boxedmm}_j \\
 {\boxedpm}_i{\boxedpp}_j & e^{2x-2x'-w} {\boxedpm}_i{\boxedpm}_j & e^{-2x+2x'-w}{\boxedpm}_i{\boxedmp}_j & {\boxedpm}_i{\boxedmm}_j \\
 {\boxedmp}_i{\boxedpp}_j & e^{-2x+2x'-w} {\boxedmp}_i{\boxedpm}_j & e^{2x-2x'-w}{\boxedmp}_i{\boxedmp}_j & {\boxedmp}_i{\boxedmm}_j \\
 e^{-2x-2x'+w} {\boxedmm}_i{\boxedpp}_j & {\boxedmm}_i{\boxedpm}_j & {\boxedmm}_i{\boxedmp}_j & e^{2x+2x'+w} {\boxedmm}_i{\boxedmm}_j \\
\end{array}
\right)
\label{TM4}
\end{eqnarray}
where $j=i+1$, $x=\beta J$, $x'=\beta J'$, $w=\beta J_{12}$, and the children's contribution functions
\begin{equation}
\boxed{{}_\pm^\pm}_i=\left[\sum_{\sigma_{i,3},\cdots,\sigma_{i,m},\cdots,\sigma_{i,M+2}}\left(e^{\beta H^{(i)}_\mathrm{children}}\right)_{{}_{\sigma_{i,1}=\pm}^{\sigma_{i,2}=\pm}}\right]^{\frac{1}{2}}.
\label{rainbow}
\end{equation}
In deriving Eqs.~(\ref{TM4}) and (\ref{rainbow}), the detailed form of $H^{(i)}_\mathrm{children}$ is not needed. It works for arbitrary forms of interactions in $H^{(i)}_\mathrm{children}$ and for both classical and quantum children, because the commutator $[H^{(i)}_\mathrm{children},H]=0$. So, the children can be more than spins. They can be electrons, phonons, excitons, Cooper pairs, fractons, anyons, etc. For the systems with mixed quantum particles and classical Ising spins~\cite{Yin_PRL_FeTe}, Eq.~(\ref{rainbow}) means that one first obtains the $2^M$ eigenvalues (energy levels) of the quantum Hamiltonian $H^{(i)}_\mathrm{children}$ for one of the four $\left({}_{\sigma_1}^{\sigma_2}\right) = \left({}_+^+\right), \left({}_+^-\right), \left({}_-^+\right), \left({}_-^-\right)$ combinations, say $\left({}_+^+\right)$, and thermally populates those energy levels to get ${\boxedpp}_i$. Then move on to work out for the other three combinations one by one.

Then, using the spin up-down symmetry in the absence of an external bias field, i.e., ${\boxedmm}_i={\boxedpp}_i$ and ${\boxedpm}_i={\boxedmp}_i$, as well as the weak condition of the minimal unit cell (i.e., no household dependence; $\boxed{{}_\pm^\pm}_i=\boxed{{}_\pm^\pm}_j=\boxed{{}_\pm^\pm}$), the transfer matrix is rewritten as
\begin{eqnarray}
\Lambda=\left(
\begin{array}{cccc}
 a & z & z & u \\
 z & b & v & z \\
 z & v & b & z \\
 u & z & z & a \\
\end{array}
\right)
\label{TM4simple}
\end{eqnarray}
where $a=e^{2x+2x'+w} \boxedpp^2, z=\boxedpp\boxedpm, u=e^{-2x-2x'+w} \boxedpp^2, b=e^{2x-2x'-w} \boxedpm^2, v=e^{-2x+2x'-w}\boxedpm^2$. It is highly symmetric or of supersymmetry, It can be block diagonalized by the parity-symmetry operations $U$ for the two spin-reversed pairs $\left({}_+^+,{}_-^-\right)$ and $\left({}_+^-,{}_-^+\right)$,
\begin{eqnarray}
U=\frac{1}{\sqrt{2}}\left(
\begin{array}{cccc}
 1 & 0 & 0 & 1 \\
 0 & 1 & 1 & 0 \\
 0 & -1 & 1 & 0 \\
 -1 & 0 & 0 & 1 \\
\end{array}
\right)
\label{U}
\end{eqnarray}
and the result is
\begin{eqnarray}
U^T \Lambda U=\left(
\begin{array}{cccc}
 a-u & 0 & 0 & 0 \\
 0 & b-v & 0 & 0 \\
 0 & 0 & b+v & 2z \\
 0 & 0 & 2z & a+u \\
\end{array}
\right)
\label{transformed}
\end{eqnarray}
where $a+u=2\cosh(2x+2x')e^w\boxedpp^2$, $b+v=2\cosh(2x-2x')e^{-w}\boxedpm^2$,
$a-u=2\sinh(2x+2x')e^w\boxedpp^2$,
$b-v=2\sinh(2x-2x')e^{-w}\boxedpm^2$, and
$z=\boxedpp\boxedpm$.
The eigensystem problem is reduced to a $2\times2$ matrix problem for the even-parity states [the bottom right part of Eq.~(\ref{transformed})], i.e.,
\begin{eqnarray}
(U^T \Lambda U)_\mathrm{even-parity}=\left(
\begin{array}{cc}
 b+v & 2z \\
  2z & a+u \\
\end{array}
\right)
\label{2x2}
\end{eqnarray}
which can be easily solved. The eigenvalues of the transfer matrix are $a-u$, $b-v$, and
\begin{equation}
\lambda_\pm=\frac{a+u+b+v}{2} \pm  \sqrt{\left(\frac{a+u-b-v}{2}\right)^2+4z^2}.
\label{general}
\end{equation}
Let us discuss two scenarios: $x'=0$ and $x'\neq 0$.

\subsection{$J'=0$}
An elegant form of the largest eigenvalue of the transfer matrix is given by
\begin{equation}
\lambda=\lambda_+=\Upsilon_+ \left[\cosh(2\beta J) + \sqrt{1+(\Upsilon_-/\Upsilon_+)^2\sinh^2(2\beta J)}\right],
\label{Z}
\end{equation}
where the frustration functions
\begin{equation}
\Upsilon_\pm=e^{\beta J_{12}}\boxedpp^2 \pm e^{-\beta J_{12}}\boxedpm^2.
\label{Y}
\end{equation}
$\Upsilon_\pm$ do not dependent on $J$ explicitly, while $\lambda$ depends explicitly on the intra-household interactions solely via $\Upsilon_\pm$.
The unconventional order parameter is the family member correlation functions~\cite{Yin_MPT}:
\begin{eqnarray}
C_{12}(0)&=&\langle \sigma_{i,1}\sigma_{i,2} \rangle_T=\displaystyle -\frac{\partial f(T)}{\partial J_{12}}=\frac{(\Upsilon_-/\Upsilon_+)\cosh(2\beta J)}{\left[{1+(\Upsilon_-/\Upsilon_+)^2\sinh^2(2\beta J)}\right]^{\frac{1}{2}}}. \label{C12}
\end{eqnarray}
where $ \langle\cdots\rangle_T $ denotes the thermodynamical average. We arrive at the same mathematical structures of Eqs.~(\ref{Z}), (\ref{Y}), and (\ref{C12}) as before~\cite{Yin_MPT}---and with the children's contribution to the frustration functions being generalized to (\ref{rainbow}).

The mathematical structure of conventional phase transitions is the non-analyticity of the system's thermodynamic free energy $f(T)$ where $T$ denotes temperature. A $k$th-order phase transition means that the $k$th derivative of $f(T)$ starts to be discontinuous at the transition. At a glance, Eq.~(\ref{Z}) and thus $f(T)$ are analytic. This is a direct consequence of the fact that the transfer matrix, made up from Boltzmann factors, i.e., exponentials, is always strictly positive, irreducible, and analytic~\cite{Cuesta_1D_PT}. However, Eq.~(\ref{Z}) has a novel mathematical structure for two features: Firstly, as the difference between two positive quantities, the frustration function \textbf{$\Upsilon_-$ can change sign} (resembling an external magnetic field) when the following condition is satisfied
\begin{equation}\label{boxed}
   \boxedpp^2 \neq  \boxedpm^2.
\end{equation}
For example, for the likely situation of $\boxedpp^2 > \boxedpm^2$ (i.e., the children contribute more when the parents hold the same values than when the parents argue), a negative $J_{12}$ decreases the impact of $\boxedpp^2$ while increasing the impact of $\boxedpm^2$. Hence, the sign of $\Upsilon_-$ can change for a suitable combination of the intra-household interactions with $J_{12}<0$. This also explain why MPT and PPPT was not found in the ordinary 2-leg ladder without the children. In that case,  $\boxedpp^2 = \boxedpm^2 = 1$, violate the condition of Eq.~(\ref{boxed}) and thus $\Upsilon_-=2\sinh(2\beta J_{12})$, which mathematically changes sign only at $\beta=0$, i.e., $T=\infty$. This could never happen.

Secondly, $(\Upsilon_-/\Upsilon_+)^2$ in Eq.~(\ref{Z}) has a prefactor of $\sinh^2(2\beta J)$, \textbf{which is exponentially large near $T_{m}$}, the temperature at which $\Upsilon_-$ changes sign. So, if Eq.~(\ref{Z}) is approximated by neglecting 1 inside $\sqrt{\cdots}$ as it is exponentially smaller than the other terms in the $\sqrt{\cdots}$,
\begin{equation}
\lambda \simeq \Upsilon_+ \cosh(2\beta J) + |\Upsilon_-|\sinh(2\beta|J|),
\label{Z2}
\end{equation}
which becomes non-analytic. This mimicking of $|\Upsilon_-|$ can also be regarded as a virtual level crossing at $T_{m}$ between $\lambda$ and the other eigenvalues of the transfer matrix, which was shown to be important for realizing phase transitions in one dimension~\cite{Cuesta_1D_PT}. The difference between Eq.~(\ref{Z}) and Eq.~(\ref{Z2}) takes place in a region of $(T_{m}-\delta T, T_{m} +\delta T)$, where $\delta T$ can be estimated by
\begin{equation}
    |\Upsilon_-/\Upsilon_+|\sinh(2\beta|J|)=1
    \label{width}
\end{equation}
at $T=T_{m}+\delta T$. $\delta T$ is proportional to $\sinh(2\beta|J|)^{-1}$ and exponentially approaches to zero (the transition is much sharper and much sharper) as $T_{m}$ decreases. Ref.~\cite{Yin_MPT} gives an example of $\delta T \approx 0.4\times 10^{-30}$~K for $T_{m}\sim 9$~K. However, for the mathematically strict $\delta T=0$ transition, $T_{m}=0$ as well~\cite{Yin_MPT}. Therefore, this kind of super-sharp transitions is named MPT. These two features must be satisfied simultaneously, to mimic the original simplest one-dimensional Ising model in an external field. That $\Upsilon_-$ changes sign at a finite temperature $T^*$ but without an exponentially large prefactor at $T^*$ is a phase crossover.

\subsection{$J' \neq 0$}
Suppose $|J'|<|J|$ without loss of generality; otherwise, we just exchange $J$ and $J'$ in all the equations presented above. In general, Eq.~(\ref{general}) tells us that the simulated non-analyticity of $|\Upsilon_-|$ takes place at $a+u-b-v=0$. When $|J'|$ and $|J|$ are substantially different, i.e., near $T_{m}$, the condition
\begin{equation}
\label{condition}
 e^{2\beta|J\pm J'|}\gg 1,
\end{equation}
is satisfied, all the aforementioned formulae for $J'=0$ are retained with only one modification:
\begin{equation}
\Upsilon_\pm=e^{\beta [J_{12}+2J'\mathrm{sgn}(J)]}\boxedpp^2 \pm e^{-\beta [J_{12}+2J'\mathrm{sgn}(J)]}\boxedpm^2.
\label{Y2}
\end{equation}
Thus, the effect of $J'$ is to make the substitution: $J_{12} \to J_{12} + 2J'\mathrm{sgn}(J)$. The condition of Eq.~(\ref{condition}) holds when $|J'|$ is appreciably different from $|J|$, which is generally true. This has two significant implications: One, MPT generally take places for $J'\neq 0$. This enlarges the parameter space of the model and stabilize the MPT against perturbations. The other is that the straightforward effect of $J'$ on $J_{12}$ provides a simple way to tune the parameters.
In the following, we will focus on the presentation on $J'=0$ and keep in mind $J_{12} \to J_{12} + 2J'\mathrm{sgn}(J)$ in general.

The unusual cases where the condition of Eq.~(\ref{condition}) does not hold can be studied with $a+u-b-v=0$ to set $T_{m}$. The special case of $M=0$, $J=J'$ and $J_{12}=-1.95|J|$ was studied in the name of an Ising tetrahedral chain composed of edge-sharing tetrahedra~\cite{Strecka_Ising_pseudo} and reinterpreted here with $T_{m}=2\alpha |J|/(k_\mathrm{B}\ln2)$, where $\alpha=(J_{12}+2|J|)/|J|$, and $\delta T/T_{m}=(2\sqrt{2}/\ln2) 2^{-1/\alpha}$. However, without the knowledge of its unconventional order parameter, the transition was categorized in the literature as one case of pseudo phase transition~\cite{Strecka_Ising_pseudo}, which deserves further comments as follows.

The pseudo phase transition was originally proposed for the pseudo critical behavior in a single chain of Ising spins periodically decorated with other particles in between~\cite{Souza_SSC_18_pPT,Rojas_PRE_99_pPT,Derzhko_pPT}, where the transfer matrix is of the following form:
\begin{eqnarray}
\Lambda_\mathrm{single-chain}=\left(
\begin{array}{cc}
  a & c \\
  c & b \\
\end{array}
\right),
\label{pPT}
\end{eqnarray}
where $a=\boxed{++}$, $b=\boxed{--}$, and $c=\boxed{+-}=\boxed{-+}$. In the absence of the external magnetic field, $a=b$ and the largest eigenvalue is $\lambda=a+c$. Then, the contribution of frustration from decoration to $a$ and $c$ may yield a peak-like feature in specific heat but not sharp enough to be considered as a pseudo critical behavior. Applying the external magnetic field makes $a \ne b$. The resultant $\lambda$ has the form consisting of a square root term like Eq.~(\ref{general}) or (\ref{Z}), or the simplest Ising model in the presence of magnetic field~\cite{Souza_SSC_18_pPT,Rojas_PRE_99_pPT,Derzhko_pPT}. As discussed above in Eqs.~(\ref{Z2}) and (\ref{width}), the square root creates a virtual level crossing experience at finite temperature, and yields the pseudo critical behavior in those decorated single-chain Ising systems~\cite{Souza_SSC_18_pPT,Rojas_PRE_99_pPT,Derzhko_pPT}. In sharp contrast, MPT is a spontaneous transition and the square root term in Eq.~(\ref{general}) or (\ref{Z}) is obtained in the absence of the external magnetic field. As a matter of fact, the MPT's response to the external magnetic field has been shown to be qualitatively different from the pseudo phase transition in the previous study of the Ising tetrahedral chain composed of edge-sharing tetrahedra, that is, the external magnetic field broadens the MPT, compared with its sharpening the pseudo phase transition~\cite{Strecka_Ising_pseudo}---however, this qualitative difference has been completely overlooked. The field-induced broadening of MPT is because the external magnetic field invalidates the reduction of the transfer matrix from $4\times 4$ to $2\times2$ for MPT (i.e., breaks the supersymmetry) and weakens the virtual level crossing experience. Therefore, I suggest that the two concepts of MPT and pseudo phase transition be clearly separated. It is this separation that leads to the present discovery of an infinite number of nontrivial MPT cases.

\section{Classification}

Since the above proof implies that we have obtained an infinite number of the MPT cases, one of the first worthy scientific activities is to classify them into distinct categories. I hereby propose a two-class classification. One is called the \emph{regular class} for the cases satisfying
\begin{equation}
J_{12} < 0  \;\;\;\mathrm{and}\;\;\;\boxedpp^2 > \boxedpm^2,
\label{regular}
\end{equation}
which means that the children contribute more when the parents hold the same values than when the parents don't. The most regular systems have $J_{1m}=J_{2m}$ for all the children $m=3,M+2$, which means that the system has the mirror symmetry in the parents-children interactions. In these systems, Eq.~(\ref{regular}) becomes apparent because the two parents having opposite values in $\boxedpm$ zero out the $J_{1m}$ and $J_{2m}$ part of the children's contribution to the total energy.

The other category is called the \emph{exotic class} for the cases satisfying
\begin{equation}
J_{12} > 0  \;\;\;\mathrm{and}\;\;\;\boxedpp^2 < \boxedpm^2,
\label{exotic}
\end{equation}
which generally lacks the mirror symmetry. Therefore, even though the two parents have opposite values in $\boxedpm$, the children's contribution to the total energy will not be zeroed out by the unequal $J_{1m}$ and $J_{2m}$.

Be alert that every regular system has two trivial exotic system companies: They are connected by the $J_{1m} \to -J_{1m}$ or $J_{2m} \to -J_{2m}$  transformation for all the children simultaneously. Thus, the first nontrivial exotic system appears for $M=2$, which breaks the mirror symmetry and retain the inversion symmetry. The existence of nontrivial exotic systems justify the generalization of the studies of MPT to increased family sizes. It also demonstrates the children's power in flipping the argumentative parents to the cooperative parents and increasing the diversity and colorfulness of MPT.

For the regular systems, the parents' direct interaction $J_{12}<0$ must be hold for the MPT to occur. The MPT is generally characterized by the order parameter $C_{12}(0)\simeq +1$ and $-1$ below $T_{m}-\delta T $ and above $T_{m}+\delta T$, respectively (exactly zero at $T_{m}$). A typical phase diagram is shown in Fig.~\ref{Fig:model}b. Although the parents tend to have unlike values, since their interactions with the children satisfy $\boxedpp^2 > \boxedpm^2$ in the low-temperature phase where energy contributions matter the most, the parents are in unison. As the system is heated up, they become less care about the family's energy need and go on to take unlike values, leaving their children in strong frustration. This leads to a large gain in the entropy's contribution to the free energy $f(T)$. Thus, the MPT is an entropy-driven first-order transition with a large latent heat, a waterfall behavior of the entropy, and a super-sharp peak in heat capacity at $T_{m}$~\cite{Yin_MPT}.

The situation is opposite for the exotic systems, where the parents' direct interaction $J_{12}>0$ must be hold for the MPT to occur. A typical phase diagram is shown in Fig.~\ref{Fig:model}c. Although the parents tend to have like values, since their interactions with the children satisfy $\boxedpp^2 < \boxedpm^2$ in the low-temperature phase where energy contributions matter the most, the parents are in disagreement. As the system is heated up, they become less care about the family's energy need and go on to take like values, being effectively detached from their children. This leads to a large gain in the entropy's contribution to the free energy $f(T)$. Thus, the MPT is also an entropy-driven first-order transition with a large latent heat, as will exemplified with a $M=2$ case below.

\section{Examples and Novel Results}
We proceed with a few examples to reveal the rich phenomena of this generalized model, including phase reentrance and the $T_{m}$ dome. We will focus on the Ising model of Eq.~(\ref{model}).

\subsection{The regular class with the mirror symmetry}
We begin with the most regular cases where $\beta J_{1m}=\beta J_{2m}=y$ and $\beta J_{mm'}=g$, i.e., all the children have the same interactions with both of their parents and the interactions between the kids are all the same. In terms of the rung geometry, they form triangles, squares (or tetrahedra), trigonal bipyramids, octahedra for $M=1,2,3,4$, respectively (Fig.~\ref{Fig:model}a excluding the rightmost one). The results for $M=1,2,3,4$ are listed in Table~\ref{table:examples}. $w<0$ must hold for the transition to take place, as discussed in the last section. $T_{m} \geq 0$ sets the other general constraint for the model parameters on achieving MPT. To have insight into how the interactions between the children affect the MPT, the results are divided into three regions: $g>0$, $g=0$, and $g<0$. They will be discussed in passing. Note that for shorthand notation, when we describe that $x$, $y$, $w$, or $g$ is strong or weak, e.g., weak $w$ (though defined as $w=\beta J_{12}$) means weak $J_{12}$, not weak $\beta J_{12}$, which is very large near $T_{m}$.

\begin{table}[b]
\begin{center}
\caption{\textbf{The regular rung-shape cases}. $M$ is the number of children per household, $\boxedpp$ and $\boxedpm$ the children's contribution functions, $T_{m}$ the transition temperature estimated from using $\Upsilon_-=0$, and $\delta T$ the transition half-width estimated from using $|\Upsilon_-/\Upsilon_+|\sinh(2|x|)=1$. The shorthand notations are $x=\beta J$, $y=\beta J_{1m} = \beta J_{2m}$, $g=\beta J_{mm'}$, and $w=\beta J_{12}<0$. The last line are the results for the exotic diamond rung (see text in the next section). $T_m/J$ means $k_\mathrm{B}T_m/J$.}
\begin{tabular}{|c|c|c|cc|cc|cc|}
\hline\hline
     &              &              & \multicolumn{2}{c|}{$g>0$}  & \multicolumn{2}{c|}{$g=0$}  & \multicolumn{2}{c|}{$g<0$}\\
 $M$ & $\boxedpp^2$ & $\boxedpm^2$ & \multicolumn{2}{c|}{$\alpha=\frac{w+M|y|}{|x|}$} &  \multicolumn{2}{c|}{$\alpha=\frac{w+M|y|}{M|x|}$} &  \multicolumn{2}{c|}{$\alpha=\frac{g+w+M|y|}{|x|}$} \\
     &              &              &  $\frac{T_{m}}{J}$ & $\frac{\delta T}{T_{m}}$ &  $\frac{T_{m}}{J}$ & $\frac{\delta T}{T_{m}}$ &  $\frac{T_{m}}{J}$ & $\frac{\delta T}{T_{m}}$\\
\hline
 1 & $2\cosh(2y)$ & 2 & $\frac{2\alpha}{\ln2}$ &$\frac{4}{\ln2} \frac{1}{2^{1/\alpha}}$ & $\frac{2\alpha}{\ln2}$ &$\frac{4}{M\ln2} \frac{1}{2^{1/\alpha}}$ & $\frac{2\alpha}{\ln2}$ &$\frac{4}{\ln2} \frac{1}{2^{1/\alpha}}$ \\
 2 & $2e^{g}\cosh(4 y) + 2e^{-g}$ & $2e^g+2e^{-g}$ & $\frac{2\alpha}{\ln2}$ & $\frac{4}{\ln2} \frac{1}{2^{1/\alpha}}$ & $\frac{2\alpha}{\ln2}$ & $\frac{4}{M\ln2} \frac{1}{2^{1/\alpha}}$& $\frac{2\alpha}{\ln2}$ &$\frac{4}{\ln2} \frac{1}{2^{1/\alpha}}$ \\
 3 & $2e^{3g}\cosh(6 y) + 6e^{-g}\cosh(2y)$ & $2e^{3g}+6e^{-g}$ & $\frac{2\alpha}{\ln2}$ & $\frac{4}{\ln2} \frac{1}{2^{1/\alpha}}$ & $\frac{2\alpha}{\ln2}$ & $\frac{4}{M\ln2} \frac{1}{2^{1/\alpha}}$ &  \multicolumn{2}{c|}{2 solutions}\\
 4 & $2e^{6g}\cosh(8 y) + 8\cosh(4y)+6e^{-2g}$ & $2e^{6g}+8+6e^{-2g}$ & $\frac{2\alpha}{\ln2}$ & $\frac{4}{\ln2} \frac{1}{2^{1/\alpha}}$ & $\frac{2\alpha}{\ln2}$ & $\frac{4}{M\ln2} \frac{1}{2^{1/\alpha}}$ & \multicolumn{2}{c|}{2 solutions} \\
 $2^*$ & $2e^{g}\cosh(2y_1+2y_2)+2e^{-g}$ & $2e^g + 2e^{-g}\cosh(2y_1-2y_2)$ & $\frac{2\alpha}{\ln2}$ & $\frac{4}{\ln2} \frac{1}{2^{1/\alpha}}$ & & & $\frac{2\alpha}{\ln2}$\footnote{$\alpha=\frac{|y_1-y_2|-w}{|x|}$ in this nontrivial exotic system} & $\frac{4}{\ln2} \frac{1}{2^{1/\alpha}}$\\
 \hline\hline
\end{tabular}
\label{table:examples}
\end{center}
\end{table}

\begin{figure}[t]
    \begin{center}
        \subfigure[][]{
            \includegraphics[width=0.48\columnwidth,clip=true,angle=0]{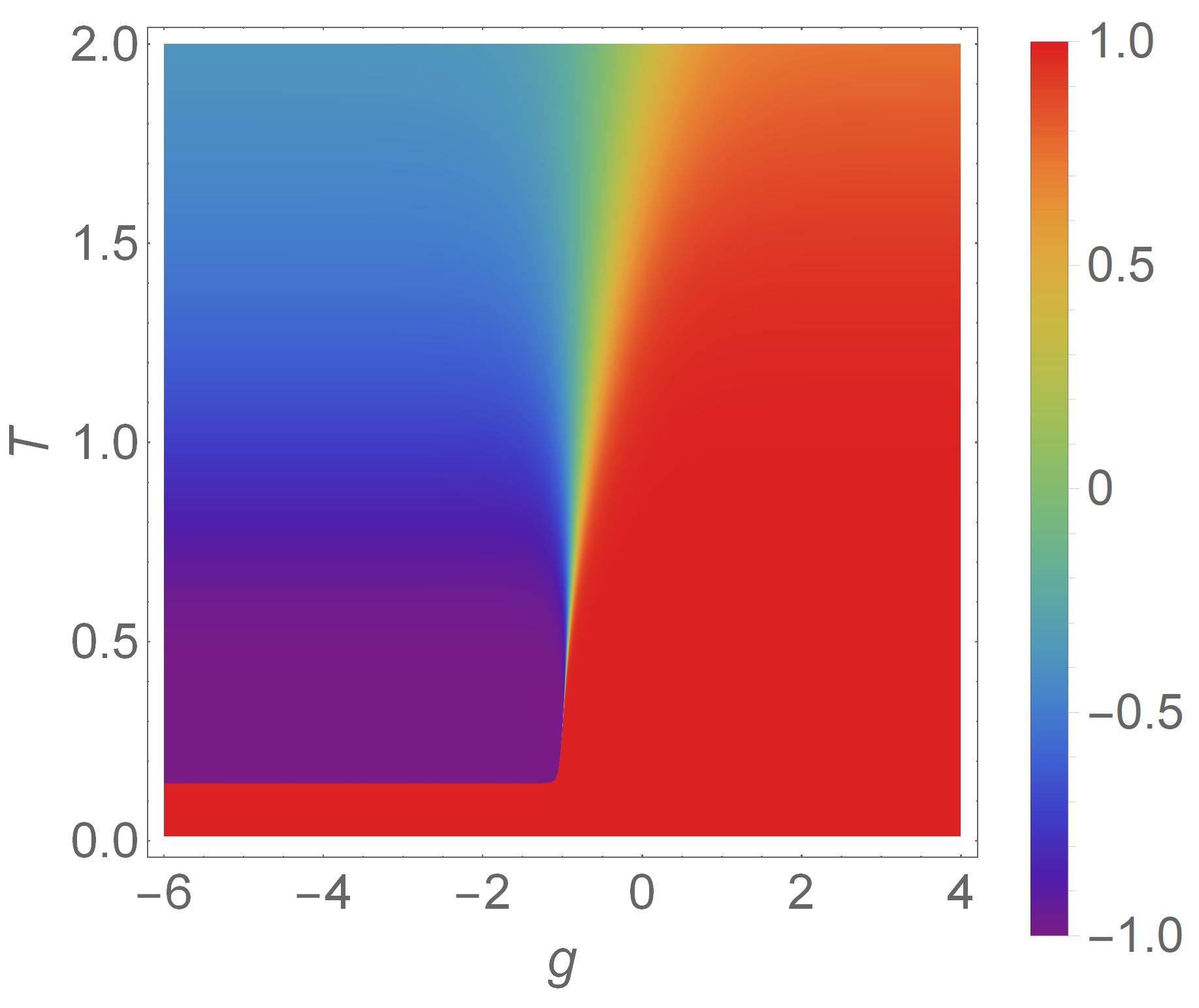}
        }
        \subfigure[][]{
            \includegraphics[width=0.48\columnwidth,clip=true,angle=0]{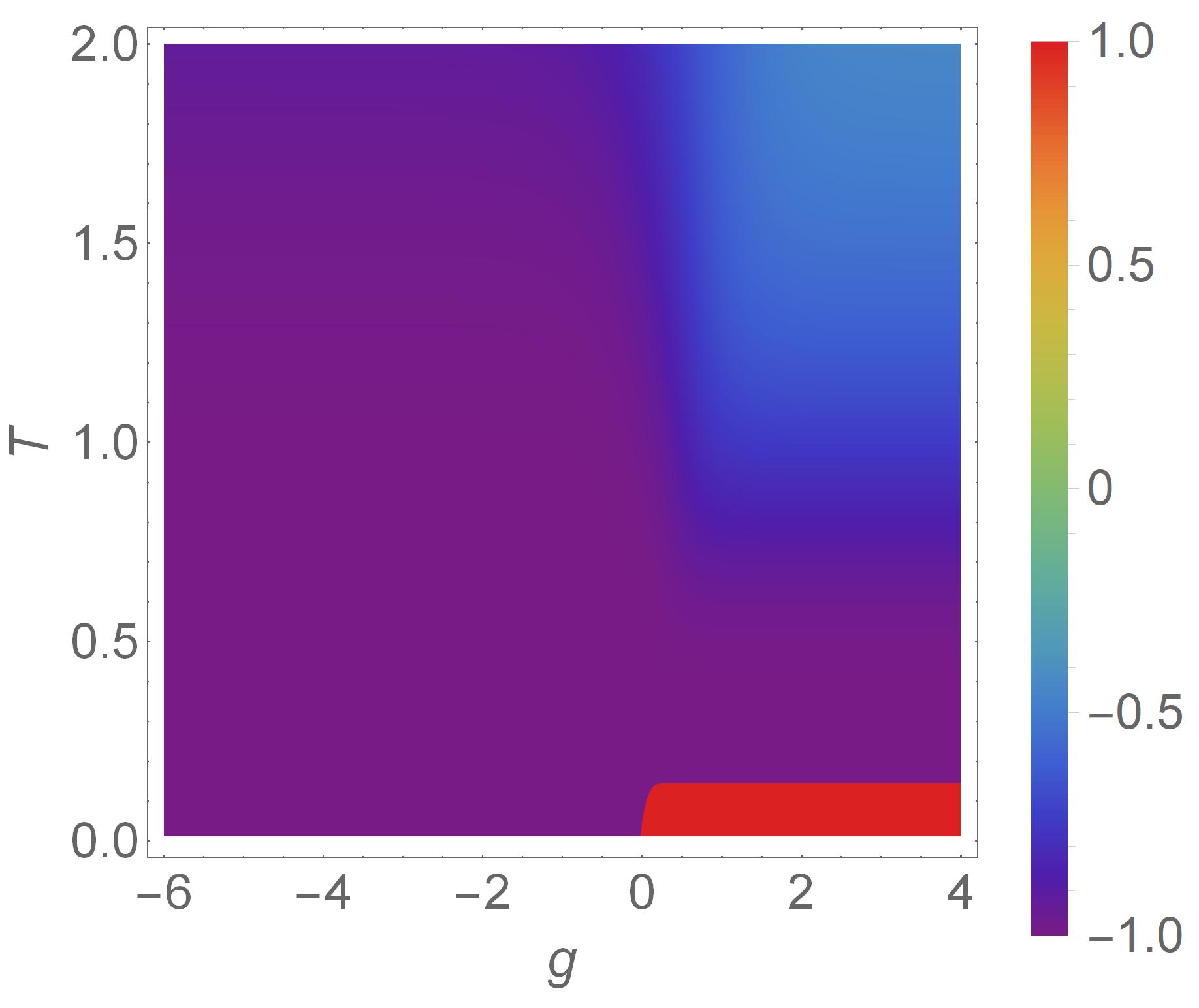}
        }        \subfigure[][]{
            \includegraphics[width=0.48\columnwidth,clip=true,angle=0]{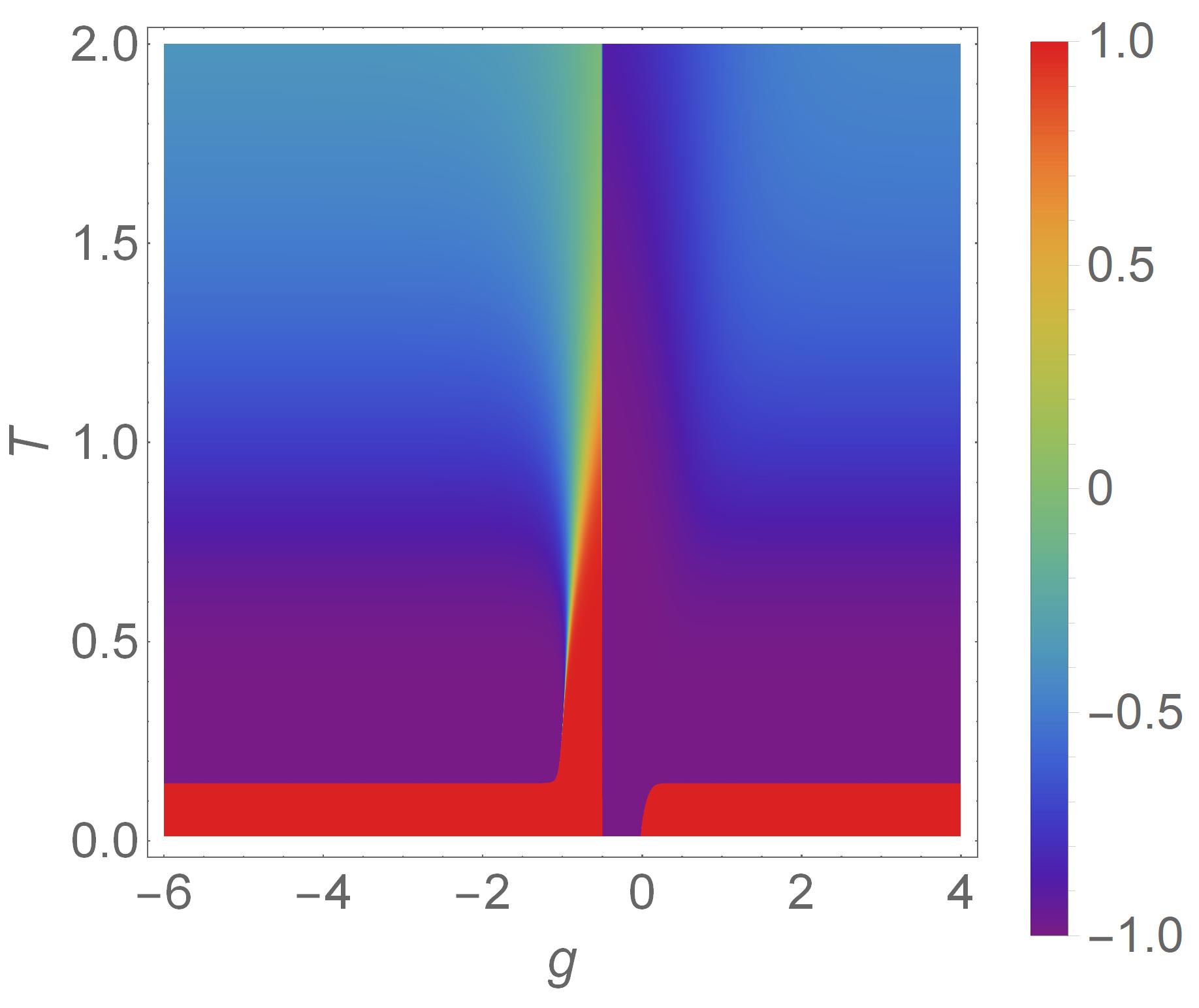}
        }
        \subfigure[][]{
            \includegraphics[width=0.48\columnwidth,clip=true,angle=0]{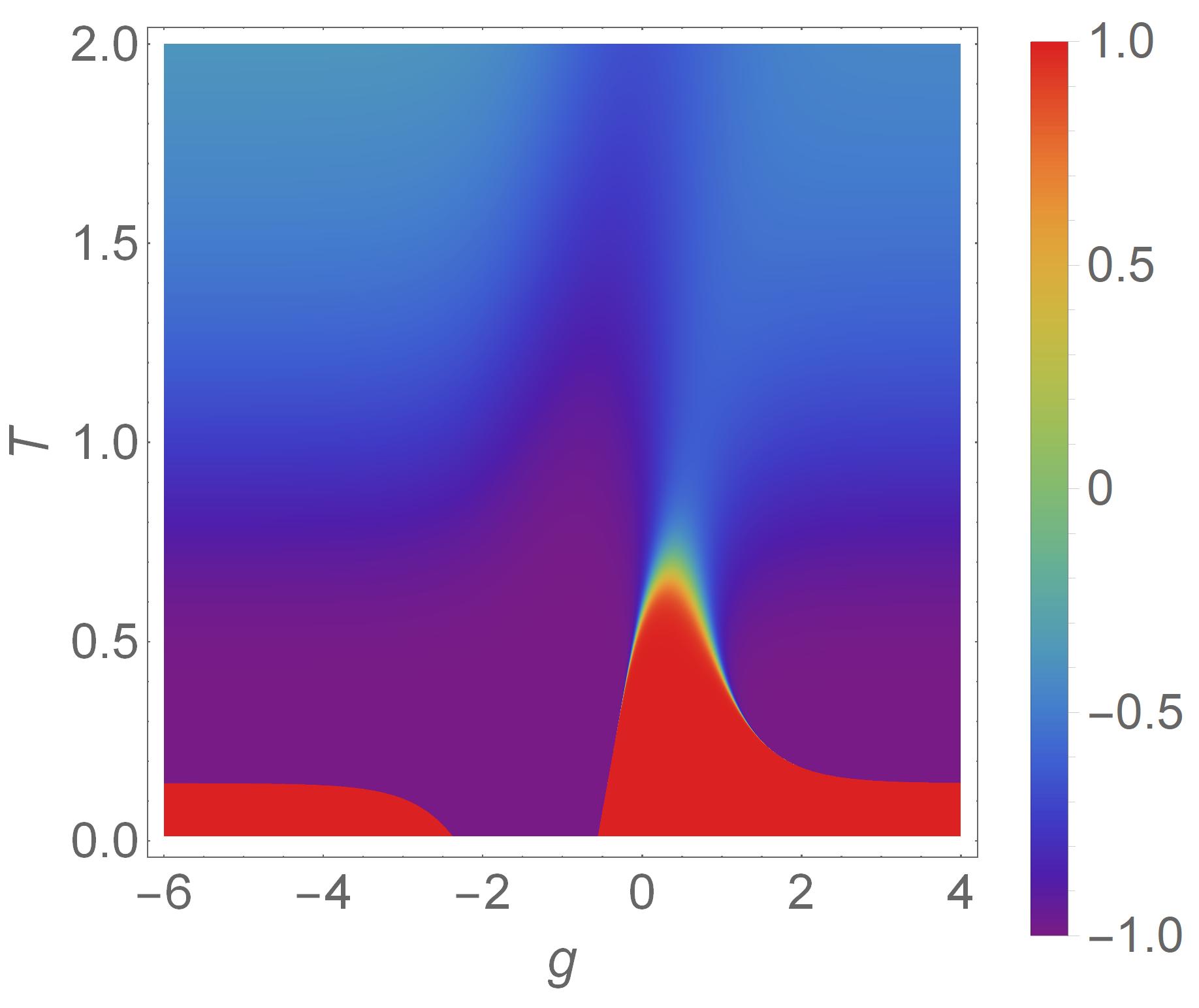}
        }
            \end{center}
\caption{\textbf{Phase diagrams of the ladder with regular trigonal bipyramid rungs} ($M=3$) in terms of the order parameter $C_{12}(0)$ as a function of $T$ the temperature and $g$ the interaction between the children. Red stands for the $+1$ region (where the parents have like values), purple stands for the $-1$ region (where the parents have unlike values). The parents' interaction $w$ is (a) $\alpha|x|-|y|$, (b) $\alpha|x|-M|y|$ for all $g$, (c) a step function of $g$ and (d) approximately a linear function of $g$ between $\alpha|x|-|y|$ and $\alpha|x|-M|y|$. The last one shows a doom-like shape of the low-temperature phase near $g=0$. Here $x$, $w$, $y_1$, $y_2$, $g$ are defined in Fig.~\ref{Fig:model}a. $J/k_\mathrm{B}$ is the temperature unit for $T$.}
\label{Fig:M3}
\end{figure}

(i) $g>0$, i.e., all the children tend to have the same value. For sufficiently strong $g$ (no need to be very large; here we show $g>0.3 J$), $\boxedpp$ and $\boxedpm$ will pick their values from the dominant first term listed in the table. The $g$ terms will be canceled out, yielding a constant $T_{m}$, which is proportional to the frustration parameter $\alpha=(w+M|y|)/|x|$. This means that the children are in unison; they act like one child with $y$ the interaction with the parents strengthened to as $M$ times as large as that in the $M=1$ case. The transition occurs at $T_{m}=2\alpha |J|/(k_\mathrm{B}\ln2)$ with the half-width $\delta T \approx 0.5$, $0.2\times 10^{-3}$, $0.1 \times 10^{-9}$, and $0.4\times 10^{-30}$~K for $\alpha=0.1, 0.05, 0.025$, and $0.01$, respectively, for a typical $|J|=300$~K (room temperature). When $g$ is positive but weak, $T_{m}$ will show a strong $g$ dependence. With the other model parameters fixed, $T_{m}$ drops by a factor of $M$ from the strong $g>0$ case to the $g=0$ case (Fig.~\ref{Fig:M3}b: the transition from the red regime to the purple one).

(ii) $g=0$, i.e., the children are neutral about influencing or being influenced by the other kids. $T_{m}$ is rescaled with $\alpha=(w+M|y|)/(M|x|)$. This means an increasing in the range of the model parameters in the case of weak $g>0$. For example, to achieve $\alpha=0.05$, $M|y|$ and $|w|$ differ by 0.05 for strong $g>0$, and by $0.05M$ for weak $g>0$. Even with the same $T_{m}$, the transition becomes sharper for weak $g$ by a factor of $1/M$ than for strong $g>0$ cases.

(iii) $g<0$, i.e., the children want to have unlike values from each other. Now, the solutions are highly $M$-dependent because with $g<0$, frustration emerges if alternating arrangement of the values cannot be satisfied simultaneous. For weak $g<0$, $\boxedpp$ still picks the first item while $\boxedpm$ picks the last item listed in the table. $\alpha$ is estimated to be $\alpha=\frac{(g+w+M|y|)}{|x|}\frac{\ln2}{\ln6}$. Thus, weak $g<0$ reduces $T_{m}$ until the phenomenon of MPT disappears. So, if changed from the regime which has MPT, (i.e., $0<\frac{w+M|y|}{|x|}<0.15$~\cite{Yin_MPT}), small $|g|/|x| \sim (w+M|y|)/|x|$ can switch off the transition (Fig.~\ref{Fig:M3}b: around $g=0$).

Moreover, starting from $M=3$, there is a strong $g<0$ solution for the MPT and PPPT. For example, for $M-3$, both $\boxedpp$ and $\boxedpm$ can pick the last items in the table, which requires the strong $g<-|y|+\frac{1}{4}\ln3$ and $\alpha=(w+|y|)/|x| > 0$ to achieve a constant $T_{m}$ for strong $g<0$ (Fig.~\ref{Fig:M3}a: the transition from the red regime to the purple one). The changing of $M|y|$ to $|y|$ in $\alpha$ and $T_{m}$ means that when the children are strongly frustrated among themselves (strong $g<0$), they no longer contribute to the transition in unison but individually. Correspondingly, for the transition to take place for strong $g<0$, the parents have to reduce the size of their disagreement parameter $w<0$ from the level of about $-M|y|$ to $-|y|$.

So, the systems need to balance the degree of frustration in order to achieve MPT. This means the intriguing \emph{reentrance behavior} of MPT. As shown in Figs.~2, how the parents adopt to their children's frustration can dramatically change the landscape of the phase diagram. In Fig.~\ref{Fig:M3}a, the parents interact with $w=\alpha|x|-|y|$ for all $g$, and the transition happens only at strong $g<0$. In Fig.~\ref{Fig:M3}b, the parents interact with $w=\alpha|x|-M|y|$ for all $g$, and the transition happens only at weak $g<0$ as a continuation of the $g>0$ solution. In Fig.~\ref{Fig:M3}c, the parents adopt a step function for adjusting $w$ to the above change in $g$. This creates two regimes with MPT separated by a regime without the transition near $g=0$. In Fig.~\ref{Fig:M3}d, the parents adopt an approximately linear  relationship in between. Most interestingly, a doom-like shape of the low-temperature phase appears. The $T_{m}$ doom is a hallmark of the phase diagrams of many strongly correlated systems. It is remarkable that we have generate something similar in this exactly solvable model with two argumentative parents and three frustrated children, although the peak area is a crossover. It is worth further studies to characterize this behavior and to reveal the $T_{m}$ dome in other more complicated cases with larger $M$.

\subsection{The exotic class with the inversion symmetry}

Here we use the simplest $M=2$ nontrivial exotic diamond rung (the rightmost one in Fig.~\ref{Fig:model}) to exemplify the essential difference from the regular class. The exotic diamond rung has inversion symmetry with $\beta J_{13}=\beta J_{24}=y_1$ and $\beta J_{14}=\beta J_{23}=y_2$ but breaks the mirror symmetry with $y_1 \neq y_2$. Suppose $y_1 > |y_2|$ without loss of generality. The results are listed in the last line of Table~\ref{table:examples} and discussed in passing.

\begin{figure}[t]
    \begin{center}
        \subfigure[][]{
\includegraphics[width=0.48\columnwidth,clip=true,angle=0]{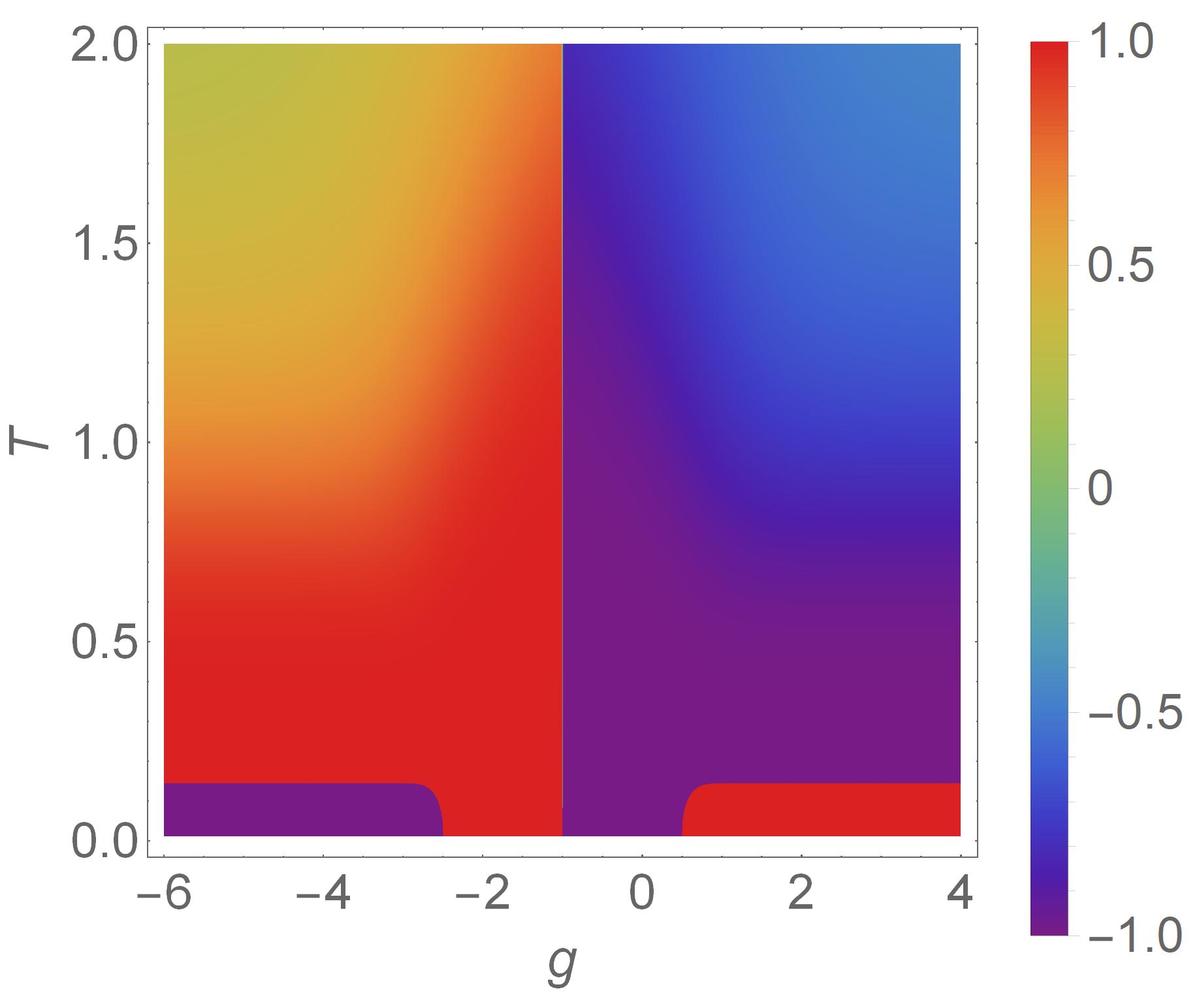}
        }
        \subfigure[][]{
\includegraphics[width=0.48\columnwidth,clip=true,angle=0]{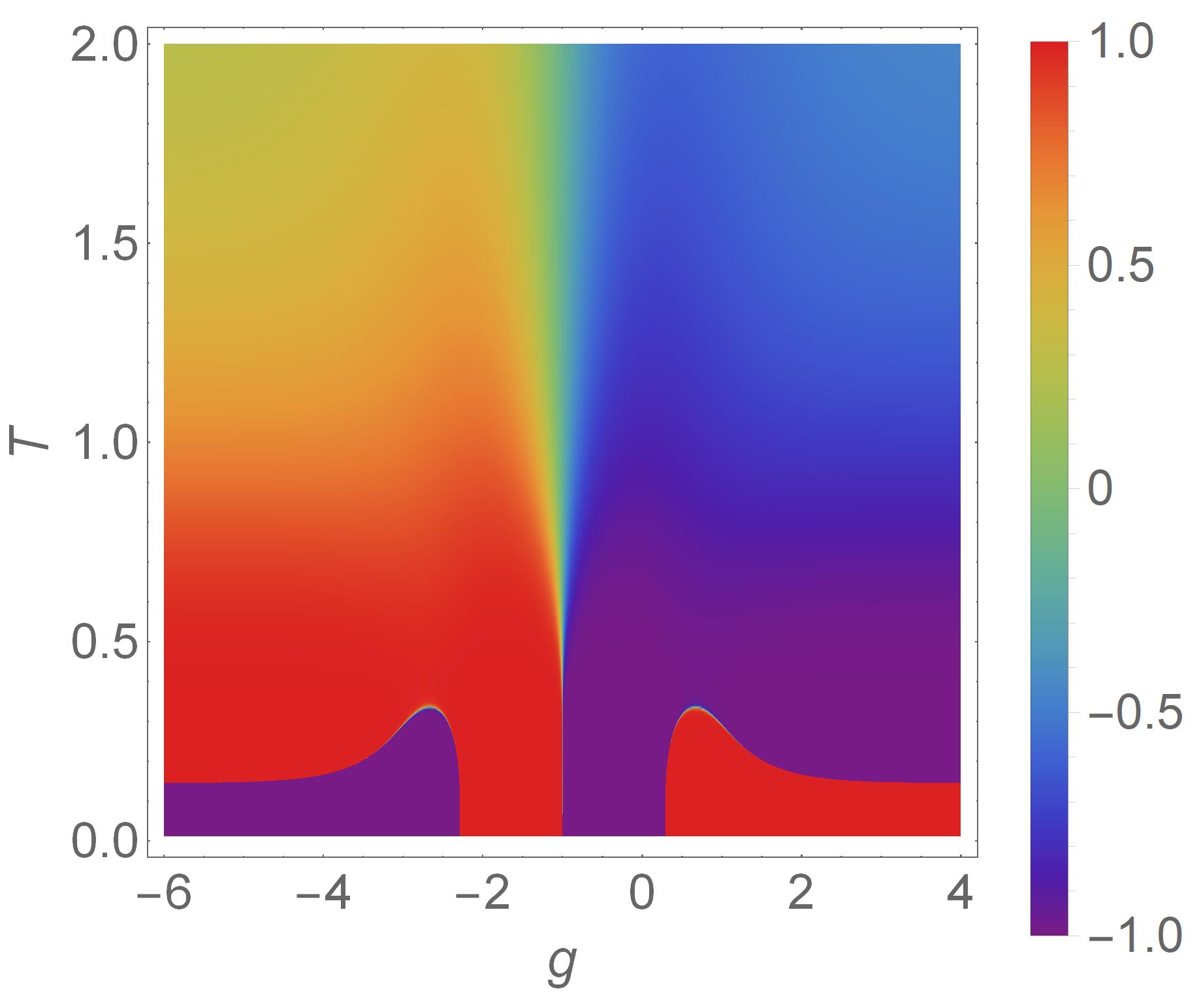}
        }
        \subfigure[][]{
\includegraphics[width=0.48\columnwidth,clip=true,angle=0]{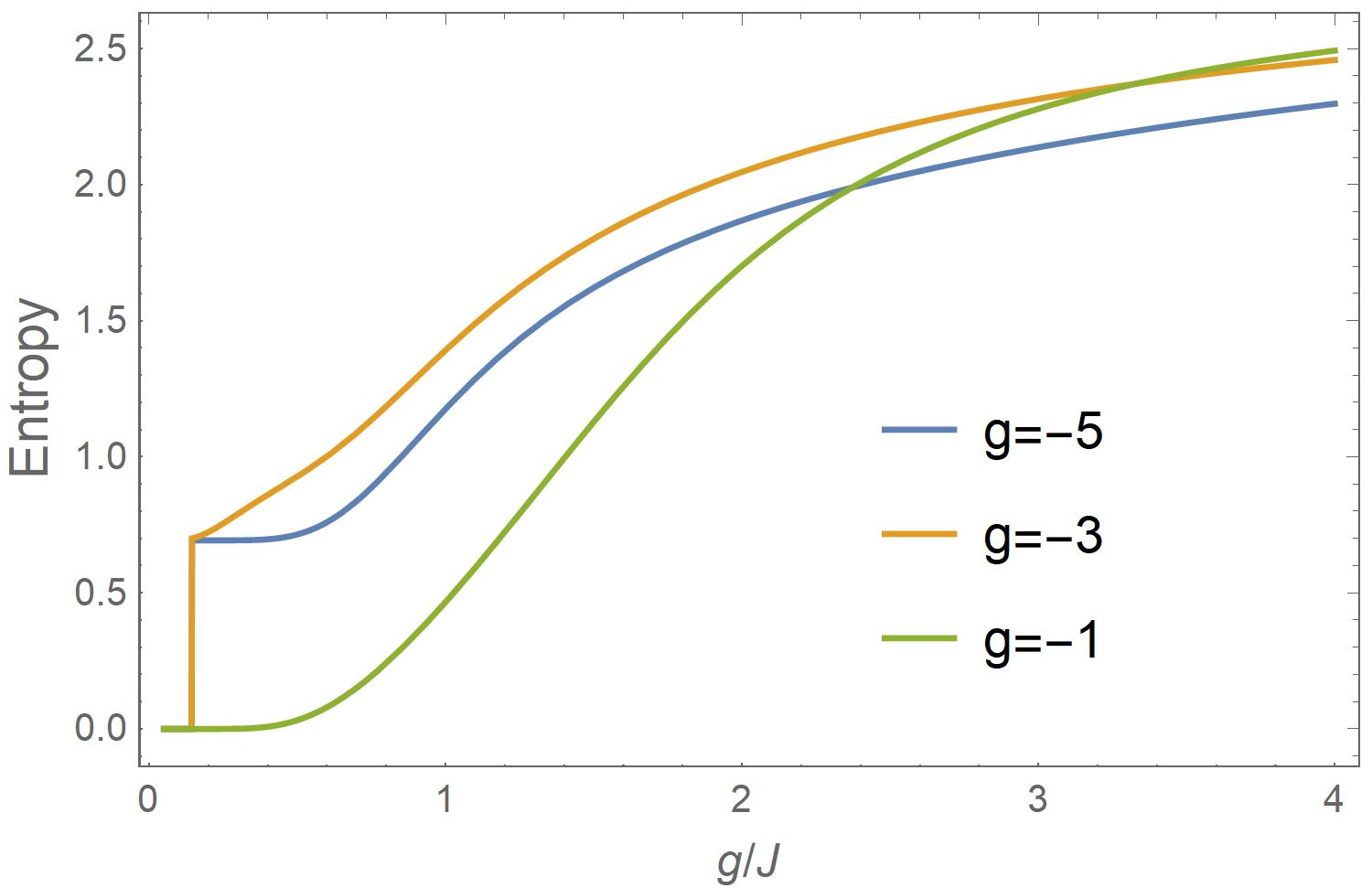}
        }
        \subfigure[][]{
\includegraphics[width=0.48\columnwidth,clip=true,angle=0]{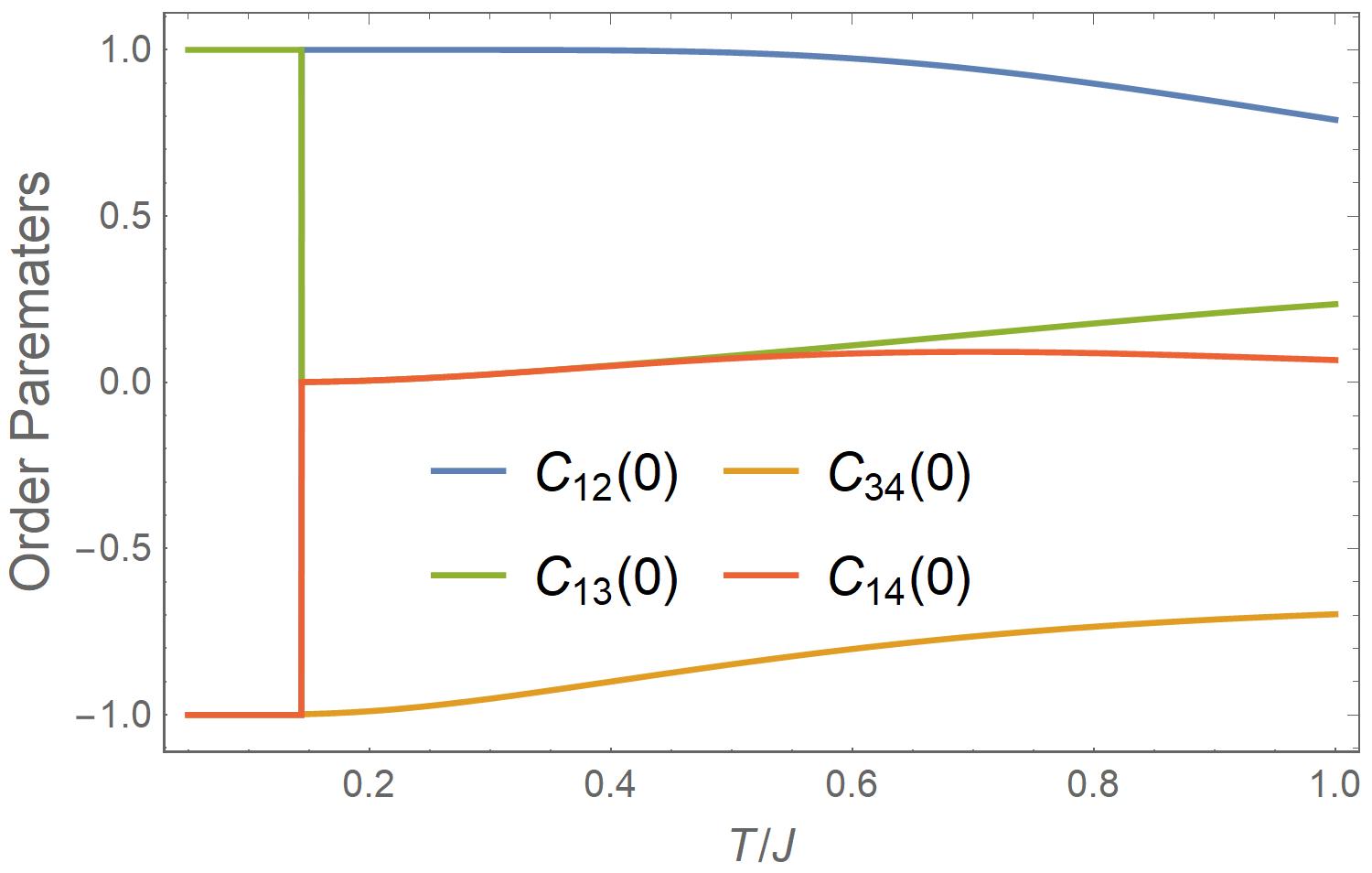}
        }
  \end{center}
\caption{\textbf{Phase diagram of the ladder with the deformed diamond rungs} in terms of the order parameter $C_{12}(0)$ as a function of $T$ the temperature and $g$ the interaction between the children. Red stands for the $+1$ region (where the parents have like values), purple stands for the $-1$ region (where the parents have unlike values). The parents' interaction $w$ is (a) a step function of $g$ and (b) approximately a linear function of $g$ between  $\alpha|x|-|y_1+y_2|$ at $g=3$ and $|y_1-y_2|-\alpha|x|$ at $g=-3$. The latter shows two hump-like shaped distinct low-temperature phases. (c) The waterfall behavior of the entropy per rung as a function of temperature for $\alpha=0.05$ (i.e., $w=0.45$, $y_1=1.55$, $y_2=1$) and three $g$'s. $|x|$ is the energy unit. (d) The same-family correlation functions $C_{mm'}(0)$ for the same set of parameters. $x$, $w$, $y_1$, $y_2$, $g$ are defined in Fig.~\ref{Fig:model}a. $J/k_\mathrm{B}$ is the temperature unit for $T$.}
\label{Fig:M2}
\end{figure}

(i) $g>0$ (the right half of Fig.~\ref{Fig:M2}a). Strong $g>0$ in fact sends the system to the regular class, because the solution for MPT simply views the system as the one with \emph{the averaged structure} with $y_{1m}=y_{2m}=(y_1+y_2)/2=y$ that recovers the mirror symmetry. Thus, $w<0$ and $\alpha=\frac{(w+M|y|)}{|x|}$. Therefore, whether a system should be classified into the regular or exotic class cannot be simply judged by the mirror symmetry between $y_{1m}$ and $y_{2m}$, which is a necessary condition for the exotic class, but not adequate.

This observation has a significant impact on one of the model's two constraints, namely minimal unit cell (i.e., no household dependence of $M$ and interactions.) While the changes in the interaction parameters can be treated continuously, the change in $M$ (and thus the structure of the interactions) is discrete and its impact can be dramatic. However, there exists at least one hopeful in the parameter space to MPT and PPPT that views the system as if it has the averaged structure.

(ii) $g<0$ (the left half of Fig.~\ref{Fig:M2}a). What is remarkably new is that for strong $g<0$, both $\boxedpp^2$ and $\boxedpm^2$ pick the last terms as listed in the table and the latter is larger now. This means that $w$ \emph{must be positive} for the transition to happen and the resulting $\alpha=(|y_1-y_2|-w)/|x|$. At a glance, this twist seems as good as the aforementioned cases. However, it brings a great benefit. In the regular cases, $\alpha=(M|y|-|w|)/|x|$. And we know that the MPT requires strong frustration, i.e., small $0<\alpha<0.15$~\cite{Yin_MPT}. This means that $|w|$ should be close to $M|y|$. Due to the geometry, the distance for $|w|$ (which is the distance between the two parents or the width of the ladder) is considerably longer than that for $|y|$ (which is the distance between a child to his/her parents). Thus, $|w| \sim M|y|$ may not be easily found in natural settings. Now, in the exotic diamond rung for strong $g<0$, $\alpha=(|y_1-y_2|-|w|)/|x|$, which means $|w| \sim |y_1-y_2|$. That is, $|w|$ is compared with the difference between two like values. This solves the problem.

As shown in Figs.~\ref{Fig:M2}a and \ref{Fig:M2}b, how the parents adopt to their children's frustration interestingly changes the landscape of the phase diagram. In Fig.~\ref{Fig:M2}a, the parents adopt a step function for adjusting $w$ to the aforementioned change in $g$. This creates two oppositely colored (regular vs exotic) regimes with MPT separated by a regime without the transition. In Fig.~\ref{Fig:M2}b, the parents adopt an approximately linear relationship in between. Two hump shapes of the low-temperature phase appear. Compared with the doom in Fig.~\ref{Fig:M3}d, the low-temperature phases live completely under the other phase, yielding the $T_{m}$ dooms.

In the exotic system ($w>0$), the parents tend to have like values. However, the parents appear to have unlike values in the ground state, because the parents have different favorite child ($y_1>y_2$) to follow and the children have unlike values ($g<0$). Above the transition temperature, the parents have the like vales. This is not something obvious to understand. In the regular system, the parents have unlike values in the higher temperature phase, and their children feel frustrated, leading to huge gain in entropy. For the exotic system, we verify that the MPT is also entropy driven, as shown in Fig.~\ref{Fig:M2}c---the waterfall behavior of the entropy for strong $g$.

To get more more insights, we examine all the same-family correlation functions
\begin{eqnarray}
C_{mm'}(0)&=&\langle \sigma_{i,m}\sigma_{i,m'} \rangle_T=\displaystyle -\frac{\partial f(T)}{\partial J_{mm'}}. \label{Cmm}
\end{eqnarray}
As shown in Fig.~\ref{Fig:M2}d, the parents' correlation function $C_{12}(0)$ changes sign at $T_{m}$, but the children's $C_{34}$ remains to be strongly negative. The correlations between a parent and all the children change to zero. That is, in the higher temperature phase, both the parent pair and the children pair enjoy their respective direct interactions and the two pairs are effectively decoupled, leading to the gain of $\ln{2}$ in entropy per household from this pairing dynamics.

\section{More general symmetry considerations}
In this section, we will discuss more symmetry related issues such as gauge freedom and the extensions of the model to include rainbow-bridge builders and supercells for MPT.

\subsection{Redundancy or gauge freedom, and its utilization}
In the derivation of the transfer matrix method, we used the \emph{weak} condition of minimum unit cell, i.e., $\boxed{{}_\pm^\pm}_i=\boxed{{}_\pm^\pm}_{i+1}$. This means that given an ice-cream-cone rung (i.e., a specific spatial distribution of the $M$ children) with $\boxed{{}_\pm^\pm}_i$, one can transform the ice cream cone with  $\boxed{{}_\pm^\pm}_i$ remaining unchanged, according to the following local (within the $i$th household) symmetry operations based on: (i) the mirror symmetry that exchange the locations of the two parents, (ii) the inversion symmetry, (iii) the mirror symmetry with respect to the bond linking the two parents, (iv) the rotational symmetry around the bond linking the two parents by an arbitrary angle. So, there are an additional infinite ways to physically distribute the children. This is particularly useful to avoid the breakdown of the system due to too crowded arrangement of the children spins,

The above symmetry-related redundancy is reminiscent of the gauge freedom associated with Berry phases and curvatures. This can have nontrivial physical effects and will be addressed in subsequent publications.

\subsection{Decorating the inter-household area}

In general, the inter-household area can also be decorated---with the participants referred to \emph{rainbow-bridge builders} (or simply \emph{bridge builders}) from now on. Then, additional interaction terms called $\sum_{i=1}^N{H^{(i,i+1)}_\mathrm{bridge}}$ are added to the model Eq.~(\ref{model}). Like the children, the bridge builders can also be integrated out. Using the spin up-down symmetry in the absence of an external bias field as well as the weak condition of the minimal unit cell, the transfer matrix is rewritten as
\begin{eqnarray}
\Lambda=\left(
\begin{array}{cccc}
 a & z & p & u \\
 z & b & v & p \\
 p & v & b & z \\
 u & p & z & a \\
\end{array}
\right)
\label{TM4general}
\end{eqnarray}
Here the shorthand notations are
\begin{eqnarray}
a&=&e^{2x+2x'+w} \boxedpp^2  \boxedpppp, \nonumber \\
u&=&e^{-2x-2x'+w} \boxedpp^2 \boxedppmm, \nonumber\\
b&=&e^{2x-2x'-w} \boxedpm^2 \boxedpmpm, \nonumber\\ v&=&e^{-2x+2x'-w}\boxedpm^2 \boxedpmmp. \\
z&=&\boxedpp\boxedpm \boxedpppm, \nonumber \\
p&=&\boxedpp\boxedpm \boxedppmp, \nonumber
\end{eqnarray}
where the boxes enclosing four signs denote the bridge builders' contribution functions
\begin{equation}
\boxed{{}_\pm^\pm{}_\pm^\pm}=\sum_{\text{bridge builders}}\left(e^{\beta H^{(i,i+1)}_\mathrm{bridge}}\right)_{
{}_{\sigma_{i,1}=\pm,}^{\sigma_{i,2}=\pm,}\;
{}_{\sigma_{i+1,1}=\pm}^{\sigma_{i+1,2}=\pm}}
\label{rainbow2}
\end{equation}
Like $H^{(i)}_\mathrm{children}$, $H^{(i,i+1)}_\mathrm{bridge}$ can be of arbitrary form, quantum or classical, and the rainbow-bridge builders can be anything, such as spins, charges, orbitons, phonons, excitons, Cooper pairs, fractons, anyons, etc. This holds as long as the parents are classical Ising spins and \textbf{there is no direct interaction between the children and the rainbow-bridge builders}. In designing, a child, say in the $i$th household, can be changed to become a rainbow-bridge builder by being connected to the parents of the $(i+1)$th household and thus being relocated into $H^{(i,i+1)}_\mathrm{bridge}$, but then the child immediately loose his/her children status---and no longer part of $H^{(i)}_\mathrm{children}$ in this model.

To make the transfer matrix approach supersymmetry like Eq.~(\ref{TM4simple}), $z=p$ is required, that is,
\begin{equation}
    \boxedpppm=\boxedppmp=\boxedmppp=\boxedpmpp.
    \label{bridge}
\end{equation}
This means that $H^{(i)}_\mathrm{bridge}$ should have either two mirror symmetries---horizontal (along the legs) and vertical---or the 4-fold rotational symmetry $C_4$ after conformational transformations (i.e., another gauge freedom), e.g., the four parents are imagined to form a square. Then, one can follow Eqs.~(\ref{U})-(\ref{general}) to exactly solve the problem to get MPT with $T_{m}$ set by $a+u-b-v=0$.

Since the bridge contribution functions included in $a$ and $u$ (or in $b$ and $v$) are generally different (i.e., $\boxedpppp \neq \boxedppmm$ or $\boxedpmpm \neq \boxedpmmp$), one will not obtain the simplest elegant form of the largest eigenvalue of the transfer matrix Eq.~(\ref{Z}), where the functionalities of the intra-household and inter-household interactions are clearly separated. It is noteworthy that while the simpler math structure does facilitate the design, it does not mean that it will be always superior for specific functionalities.

\ignore{Thus, we suggest to reserve the nickname of PPPT to those with Eq.~(\ref{Z}) and $T_{m}$ being set by the frustration function $\Upsilon_-=0$ (i.e., adding children only, which can be done arbitrarily). The more general cases with $T_{m}$ set by $a+u-b-v=0$ will be called MPT. Apparently, PPPT with an infinite number of cases is a subset of MPT. However, this suggestion comes from pure aesthetic consideration. While the simpler math structure of PPPT does facilitate the design, it does not mean that it will be always superior for specific functionalities.}

One of the simplest cases of decoration with bridge builders is adding one spin of Ising type at the center of the inter-household area, which interacts with the four surrounding parents equally by superexchange $J_b$. For $J'=0$ and $J_{12}<0$, one obtains $T_{m}=2\alpha |J|/(k_\mathrm{B}\ln2)$ where $\alpha=(J_{12}+2|J_b|)/|J|$ and $\delta T/T_{m}=(2\sqrt{2}/\ln2)2^{-1/\alpha}$. The transition is even sharper than most cases listed in Table~\ref{table:examples}.

\subsection{Supercell: beyond minimal unit cell}
The above derivation for the bridge spins implies a straightforward application---to extend the minimal unit cell ($L=1$) to the supercell with $L>1$ in which the households in between the two edge households of the supercell are regarded as bridge builders. For the supercell household issues without additional bridge builders, one horizontal (along the legs) mirror symmetry between the two edge households of the supercell is sufficient to meet the above derivation. Hence, suppose the edge household of the supercell is called $A$ and the households in between are called $B$, $C$, $\ldots$, etc. The patterns satisfying Eq.~(\ref{bridge}) are
\begin{eqnarray*}
L&=&1,\;\; \cdots AAAAAAAAAAAAA\cdots \nonumber\\
L&=&2,\;\; \cdots ABABABABABABA\cdots \nonumber\\
L&=&3,\;\; \cdots ABBABBABBABBA\cdots \nonumber\\
L&=&4,\;\; \cdots ABBBABBBABBBA\cdots \nonumber\\
L&=&4,\;\; \cdots ABCBABCBABCBA\cdots \nonumber
\end{eqnarray*}
and so on.

\section{Implications: A start toward the Ising Machine}

I have shown that by exploring the supersymmetry of the new mathematical structure 
that has not appeared before in phase-transition problems, I discovered an infinite number of cases of MPT and highly tunable exotic properties such as phase reentrance, $T_{m}$ domes, pairing, and gauge freedom, which are often seen in many other strongly correlated systems. This raises the hope that the complexity in strongly correlated electronic systems can be simulated by a MPT-based one-dimensional Ising Machine, which can be an abstract (software) or real (hardware) machine and has an infinite number of tunable inputs in design and control. In principle, this possibility is in harmony with the spirit of the Hubbard-Stratonovich transformation, which envisioned that strong quantum correlation can be simulated with the aid of  fluctuating classic field~\cite{Hubbard_PRL_59}. The present discoveries mark a start of the journey toward the Ising Machine.

One key benefit of the one-dimensional Ising Machine is that it can provide exact or highly accurate non-perturbative beyond-mean-field solutions. Such insights are unambiguous, no hand-waving, and may guide the synthesis of next-generation functional materials and the making of the first-generation phase-transition-ready one-dimensional devices for temperature sensitive applications. They may also guide the development of advanced correct theory for frustration---either geometrical or quantum---driven phase transitions in strongly correlated systems.

To build the Ising Machine, we start with the ordinary 2-leg ladder~\cite{Dagotto_science_ladder} made up by the parent spins of Ising type. Then, we decorate the parent backbone by adding/rearranging children to the households (rungs) or adding/rearranging bridge builders to the inter-household area. The children and bridge builders can be classical or quantum, can be spins, electrons, phonons, excitons, fractons, anyons, or Cooper pairs, etc. And there are additional gauge freedoms to use. Adding children can be done arbitrarily, and add bridge builders according to the symmetry of Eq.~(\ref{bridge}). Note that this does not mean the nonexistence of MPT in the cases violating Eq.~(\ref{bridge}); we only show that the cases satisfying Eq.~(\ref{bridge}) have been proven to exhibit MPT. Likewise, while adding children only has the advantages of being arbitrarily doable and having the simplest mathematical structure---thus facilitating the design, this does not mean that it will be superior to adding bridge builders for specific functionalities, as exemplified in Section VI. Designing and optimizing either children or bridge builders or both will turn out to be an endless intellectual and engineering game. As a special suggestion for the design patterns, the exotic diamond rung, which has been shown above to solve a practical issue, is conformationally identical to a Feynman diagram. Feynman diagram is an elegant and intuitive representation of quantum-field-theory processes and interactions between particles. One may build the Ising Machines in which the decorations are done based on the Feynman diagrams.

The present results have resolved a number of fundamental and practical issues concerning the applications of MPT that require the systems to be stable in performance, tunable in functionality, and sustainable in R\&D.
Since the Ising model has already been implemented in electronic circuits~\cite{Ising_FPGA} and designed for optical neural networks~\cite{Inagak_IsingMachine_Science16,Inagak_IsingMachine_Science16,Pierangeli_IsingMachine_PRL19,Yamamoto_npjQI_Ising}, even the hardware-type MPT-based one-dimensional Ising Machine appears to be immediately feasible. Considering similar hardware readiness for the quantum counterpart of the Ising model, namely the Heisenberg model, achieved during the quest for universal quantum computation using Heisenberg exchange interactions~\cite{DiVincenzo_Nature_QuantumComputation,Arute_quantum_supremacy}, it is interesting to explore unconventional phase transitions in one-dimensional Heisenberg models based on the present pool of an infinite number of MPT cases.

\begin{acknowledgments}
The author is grateful to Fang Dong for stimulating discussions. Brookhaven National Laboratory was supported by U.S. Department of Energy (DOE) Office of Basic Energy Sciences (BES) Division of Materials Sciences and Engineering under contract No. DE-SC0012704.
\end{acknowledgments}

\bibliography{Ising_quantum}



\newpage

\newpage

\newpage

\end{document}